\documentclass[12pt,twoside,a4paper]{article}
\usepackage{tikz}

\usepackage{geometry,color,framed}
\usepackage{amsmath,amssymb,mathtools,slashed,xcolor,mdframed}
\usepackage{amstext}
\usepackage{graphicx}
\usepackage{verbatim}
\usepackage{bm}% bold math
\usepackage[nottoc,notlot,notlof]{tocbibind}
\usepackage{caption, subcaption}
\usepackage[colorlinks=false, linkcolor= blue,linktoc=true]{hyperref}
\makeatletter

\newcommand{\beq}{\begin{eqnarray}}
\newcommand{\eeq}{\end{eqnarray}}
\newcommand{\non}{\nonumber\\}
\newcommand{\ben}{\begin{enumerate}}
\newcommand{\enn}{\end{enumerate}}
\newcommand{\lf}{\left}
\newcommand{\rr}{\right}
\newcommand{\mc}{\mathcal}
\newcommand{\bei}{\begin{itemize}}
\newcommand{\eni}{\end{itemize}}
\newcommand{\tb}{\textbf}
\newcommand{\ttx}{\texttt}
\newcommand{\ra}{\rightarrow}
\newcommand{\mb}{\mathbb}
\newcommand{\mf}{\mathbf}
\def\pd{{\partial}}
\def\m{{\mu}}
\def\n{{\nu}}

\def\d{{\delta}}

\def\p{{\phi}}

\def\S{{\Sigma}}

\def\O{{\Omega}}
\input{epsf.sty}  
\begin{document}

\begin{center}
\[\]\[\]\[\]
\LARGE{Cosmological solutions from 5D \\$N=4$ matter-coupled supergravity}
\end{center}
\vspace{1cm}
\begin{center}
\large{H. L. Dao}
\end{center}
\begin{center}
\textsl{Department of Physics,\\
 National University of Singapore,\\
3 Science Drive 2, Singapore 117551}\\
\texttt{hl.dao@nus.edu.sg}\\
\vspace{2cm}
\end{center}
\begin{abstract}
From five-dimensional $N=4$ matter-coupled gauged supergravity,  smooth time-dependent cosmological solutions, connecting a $dS_{5-d}\times H^d$ (with $d=2,3$) spacetime at early times to a $dS_5$ spacetime at late times, are presented. The solutions are derived from the second-order equations of motion arising from all the gauged theories that can admit $dS_5$ solutions. There are eight such theories constructed from  gauge groups of the form $SO(1,1)\times G_{nc}$ and $SO(1,1)^{(n)}_\text{diag}\times G_{nc}$, with $n=2,3$, where $G_{nc}$ is a non-compact gauge factor whose compact part must be embedded entirely in the matter symmetry group of 5D matter-coupled supergravity. Furthermore, we analyze how the cosmological solutions and their corresponding $dS_5$ vacua cannot arise from the first-order equations that solve the second-order field equations. 
\end{abstract}
\newpage
\setcounter{tocdepth}{2}
\tableofcontents
% \newpage
%%%%%%%%%%%%%%%%%%%%%%%%%%%%%%%%%%%%%%%%%%%%%%%%%%%%%%%%%%%%%%%%%
\section{Introduction}
%%%%%%%%%%%%%%%%%%%%%%%%%%%%%%%%%%%%%%%%%%%%%%%%%%%%%%%%%%%%%%%%%
The dS/CFT correspondence, as formulated in \cite{Strominger-01}, \cite{Strominger-01-2}, conjectures that the universe is a cosmological renormalization group (RG) flow between a de-Sitter spacetime in the infinite past with a very large Hubble constant (of order $10^{24}$ cm$^{-1}$) and another de-Sitter spacetime in the infinite future with a much smaller Hubble constant (of order $10^{-28}$ cm$^{-1}$). The ratio between these Hubble constant is in the order of $10^{52}$ \cite{Strominger-01-2}. Although realistic cosmological solutions arising from string theory should reproduce this observation, the lack of a stable de-Sitter solution in string theory has been a long-standing problem that hampers progress in this direction \cite{Danielsson-18}. 
\\\indent
Stable classical de-Sitter solutions in gauged supergravity would be useful in the pursuit of an eventual formulation of dS/CFT. In this same context, intra-dimensional cosmological solutions of a $D$-dimensional supergravity connecting a $dS_D$ spacetime in the infinite past to another $dS_D$ spacetime in the infinite future, or inter-dimensional cosmological solutions interpolating between some $dS_{D-d}\times \S_d$ (where $\S_d = S^d$ or $H^d$) spacetime at early times to some $dS_D$ spacetime at late times, would be the fundamental and essential ingredients of dS/CFT. 
Although dS/CFT holography is a compelling motivation for the study of $dS$ and cosmological solutions, it must be noted that the status of dS/CFT is currently far from settled. Pehaps the most promising approach to dS/CFT, at present, is one that makes use of higher spin gravity instead of conventional supergravities \cite{Strominger-11},  \cite{Anninos-14}, \cite{Anninos-17}, \cite{Neiman-18}. 
Furthermore, conventional gauged supergravities are rife with unstable $dS$ vacua, which are understood not to be useful in the context of holography. Consequently, cosmological solutions associated with these unstable $dS$ solutions might be deemed  irrelevant for the ultimate application of dS/CFT holography. 
\\\indent
Nevertheless, it can still be instructive to study $dS$ vacua and cosmological solutions in gauged supergravities in their own right, independently of any contexts related to holography. At the very least, these solutions can offer insights into the dS vacuum structure of gauged supergravities and extend the catalog of known cosmological solutions, which only contains a few examples to date. Some of the most well-known cosmological solutions, such as \cite{Lu-03}, are found in the de-Sitter supergravities that are obtained by dimensional reduction \cite{Liu-Sabra-03} from the exotic $\star$-theories that are themselves related to the conventional M and type IIB theories by timelike T-dualities  \cite{Hull-98-Jun}, \cite{Hull-98-Jul}, \cite{Hull-98-Aug}, \cite{Hull-99-time}. As a result,   de-Sitter supergravity theories have the wrong signs for the kinetic terms of the gauge field strengths - a feature which makes these theories pathological. Furthermore, these supergravities have non-compact R-symmetry groups and hence naturally admit de-Sitter spacetime as solutions.   In the context of these exotic $D$-dimensional supergravity theories, it is very natural to study $dS_D$ vacua and their associated cosmological solutions that interpolate between a $dS_{D-d}\times \S_d$ spacetime in the infinite past and a $dS_D$ spacetime in the infinite future. These cosmological solutions are often derived from second-order equations of motion, but they can also arise from some first-order system which solves the aforementioned second-order system. The existence of this first-order system is largely due to the setting of dS supergravities which favor $dS$ solutions. 
This is completely analogous to the fact that conventional $D$-dimensional supergravity theories with compact R-symmetry group readily admit fully supersymmetric $AdS_D$ as their solutions, and also RG flows interpolating between these $AdS_D$ and some $AdS_{D-d}\times \S_d$ solutions. The RG flows are solutions of the first-order BPS equations obtained by setting to zero the supersymmetric transformations of the fermionic fields in the theory. 
\\\indent
Cosmological solutions arising from a first-order system in a conventional supergravity with gauged non-compact R-symmetry  can be found in \cite{Cvetic-noncompact}. The flow solution described there runs between a singularity (possibly of the Big Bang type) to a $dS$ critical point, due to the lack of another $dS$ extremum. In \cite{Lu-03-flow}, a cosmological solution interpolating between $dS_2\times S^{D-2}$ in the infinite past and a $dS_D$ spacetime in the infinite future was obtained in $D$-dimensional de-Sitter Einstein-Maxwell gravity (with standard sign for the kinectic term of the gauge field strength) that can be embedded in conventional $M$-theory \cite{Lu-04}.  This type of solutions, obtained by analytically continuing the  supersymmetric $AdS_{D-2}\times \Omega^2$ solutions (where $\O^2 = S^2$ or $H^2$), can also arise from a system of first-order equations.
\\\indent
In this work, we perform a systematic study of cosmological solutions in the context of 5D half-maximal matter-coupled supergravity. The solutions studied here, obtained by solving the second order equations of motion, smoothly interpolate between a $dS_{5-d}\times \S_d$ ($d=2,3$) spacetime in the infinite past to a $dS_5$ in the infinite future. The $dS_5$ vacuum is a solution of 5D $N=4$ matter-coupled supergravity with some very specific gauge groups that were derived and classified in \cite{dS5} using the embedding tensor formalism \cite{Schon-Weidner}.  Note that the eight gaugings derived in \cite{dS5} and studied here completely exhaust the list of all possible semisimple gauge groups capable of admitting $dS_5$ vacua. 
Our prime motivation for this study is to determine whether there exist cosmological solutions in 5D $N=4$ supergravity given the $dS_5$ vacua found in \cite{dS5}.  Being unstable, these vacua are not meant for the application of dS/CFT, but simply as a means to gain insights into the $dS$ vacuum structure of 5D $N=4$ gauged supergravities. 
\\\indent
The organization of this paper is as follows. In section \ref{N4_SUGRA}, we review the five-dimensional $N=4$ supergravity and in section \ref{dS5-rev}, we review all the gauge groups that can give rise to $dS_5$ solutions in $N=4$ supergravity. In section \ref{ans}, we specify the ansatze and derive the second order equations of motion for the cosmological  solutions of $N=4$ 5D supergravity. In section \ref{all-G}, we present all solutions, and in section \ref{pseudo}, we analyze how these solutions cannot be derived from the first-order system of equations that solve the second-order field equations. Section \ref{concl} concludes the paper. 
%%%%%%%%%%%%%%%%%%%%%%%%%%%%%%%%%%%%%%%%%%%%%%%%%%%%%%%%%%%%%%%%%
\section{5D $N=4$ matter-coupled gauged supergravity}\label{N4_SUGRA} 
%%%%%%%%%%%%%%%%%%%%%%%%%%%%%%%%%%%%%%%%%%%%%%%%%%%%%%%%%%%%%%%%%
In this section, we briefly review five dimensional $N=4$ gauged supergravity coupled to $n$ vector multiplets. We will only highlight some basic features of the theory to make this article self-contained. The detailed construction of the theory can be found in \cite{Schon-Weidner} and \cite{5D_N4_Dallagata}. 
\\\indent
The field content of five-dimensional $N=4$ supergravity contains one supergravity multiplet and an arbitrary number $n$ of vector multiplets. The supergravity multiplet 
\beq
\lf(e^{\hat{\mu}}_\mu,\,\psi_{\mu i},\, \chi_i,\,A^0,\,A^m,\,\Sigma\rr) \nonumber
\eeq
consists of the graviton $e^{\hat{\mu}}_\mu$, four gravitini $\psi_{\mu i}$, six vectors $A^0$ and
$A_\mu^m$, four spin-$\frac{1}{2}$ fields $\chi_i$ and the dilaton
 $\Sigma$. Space-time and tangent space indices are denoted respectively by $\mu,\nu,\ldots =0,1,2,3,4$ and
$\hat{\mu},\hat{\nu},\ldots=0,1,2,3,4$. The $SO(5)\sim USp(4)$
R-symmetry indices are labeled by $m,n=1,\ldots, 5$ for the
$SO(5)$ vector representation and $i,j=1,2,3,4$ for the $SO(5)$
spinor or $USp(4)$ fundamental representation.
 The vector multiplet
 \beq
 \lf(A_\mu,\, \lambda_i, \,\phi^m\rr)
 \nonumber
 \eeq
  consists of a vector field $A_\mu$, four gaugini $\lambda_i$ and five scalars $\phi^m$. The $n$ vector multiplets will be labeled by indices $a,b=1,\ldots, n$, and the component fields within these multiplets will be denoted by 
\beq
(A^a_\mu,\lambda^{a}_i,\phi^{ma})\,\,.\nonumber
\eeq 
In total there are  $6+n$ vector fields, from the supergravity and vector multiplets, which will be collectively denoted by $A^{\mc{M}}_\mu=(A^0_\mu,A^M_\mu)=(A^0_\mu,A^m_\mu,A^a_\mu)$. 
\\\indent
The dilaton $\S$ and the $5n$ scalar fields from the vector multiplets parametrize the following scalar manifold
\beq
\mc M_\text{scalar} = SO(1,1)\times \frac{SO(5,n)}{SO(5)\times SO(n)}. \label{sm}
\eeq
The $SO(1,1)\sim \mathbb{R}^+$ factor of (\ref{sm}) is spanned by the dilaton $\S$, while the second factor $SO(5,n)/SO(5)\times SO(n)$ is spanned by scalars $\phi^{ma}$ from the vector multiplets. 
This coset manifold is described by a coset representative $\mc{V}_M^{\phantom{M}A}$ transforming under the global $SO(5,n)$ and the local $SO(5)\times SO(n)$ by left and right multiplications, respectively. Indices $M,N,\ldots=1,2,\ldots , 5+n$ are used to label the fundamental representation of the global $SO(5,n)$ group, while local $SO(5)\times SO(n)$ indices $A,B,\ldots$ will be split into $A=(m,a)$. Accordingly, the coset representative can be written as
\begin{equation}
\mc{V}_M^{\phantom{M}A}=(\mc{V}_M^{\phantom{M}m},\mc{V}_M^{\phantom{M}a}).
\end{equation}
The matrix $\mc{V}_M^{\phantom{M}A}$ is an element of $SO(5,n)$ and satisfies the relation
\begin{equation}
\eta_{MN}={\mc{V}_M}^A{\mc{V}_N}^B\eta_{AB}=-\mc{V}_M^{\phantom{M}m}\mc{V}_N^{\phantom{M}m}+\mc{V}_M^{\phantom{M}a}\mc{V}_N^{\phantom{M}a}
\end{equation}
with $\eta_{MN}=\textrm{diag}(-1,-1,-1,-1,-1,1,\ldots,1)$ being the $SO(5,n)$ invariant tensor. Note that $SO(5,n)$ indices $M,N,\ldots$ are lowered and raised by $\eta_{MN}$ and its inverse $\eta^{MN}$, respectively. 
\\
\indent Gaugings are performed by selecting a subgroup $G_0$ of the full global symmetry $G=SO(1,1)\times SO(5,n)$ to be a local symmetry. The gauging procedure is effectively described in the embedding tensor formalism. For five-dimensional $N=4$ supersymmetry, the embedding tensor has three components $\xi^{M}$, $\xi^{MN}$ and $f_{MNP}$ \cite{Schon-Weidner}.  These introduce the following minimal coupling of various fields via the covariant derivative
\begin{equation}
D_\mu=\nabla_\mu -A^M_\mu{f_M}^{NP}t_{NP}-A^0_\mu \xi^{NP}t_{NP}-A^M_\mu \xi^Nt_{MN}-A^M_\mu\xi_Mt_0
\end{equation}
in which $\nabla_\mu$ is the usual space-time covariant derivative and $t_{MN}=t_{[MN]}$ and $t_0$ are generators of $SO(5,n)$ and $SO(1,1)$, respectively. It should also be noted that $\xi^M$, $\xi^{MN}$ and $f_{MNP}$ include the gauge coupling constants.
 In terms of the embedding tensor, gauge generators ${X_{\mc{M}\mc{N}}}^{\mc{P}}={(X_{\mc{M}})_{\mc{N}}}^{\mc{P}}$ are given by
\begin{eqnarray}
{X_{MN}}^P&=&-{f_{MN}}^P-\frac{1}{2}\eta_{MN}\xi^P+\delta^P_{[M}\xi_{N]},\nonumber \\
{X_{M0}}^0&=&\xi_M,\qquad {X_{0M}}^N=-{\xi_M}^N\, .\label{gauge-gen}
\end{eqnarray}
These generators must form a closed subalgebra of $G$ with the commutation relations
\begin{equation}
\left[X_{\mc{M}},X_{\mc{N}}\right]=-{X_{\mc{M}\mc{N}}}^{\mc{P}}X_{\mc{P}},
\end{equation} 
which are enforced by the following quadratic constraints on the embedding tensor components
\begin{eqnarray}
\xi^M\xi_M&=&0,\qquad \xi_{MN}\xi^N=0,\qquad f_{MNP}\xi^P=0,\nonumber\\
3f_{R[MN}{f_{PQ]}}^R&=&2f_{[MNP}\xi_{Q]},\qquad {\xi_M}^Qf_{QNP}=\xi_M\xi_{NP}-\xi_{[N}\xi_{P]M}\, .\label{QC}
\end{eqnarray}
\\
\indent 
In this work, we are interested in cosmological solutions whose ansatze require only the metric, the dilaton $\S$ and some gauge fields $A^{\mc M}_{\m}$. Accordingly, the bosonic Lagrangian of interest to us reads
\beq
e^{-1}\mc{L}&=&\frac{1}{2}R -\frac{1}{4}\S^2 M_{MN}\mc H_{\m\n}^M \mc H^{N\m\n} -\frac{1}{4} \S^{-4}\mc H^0_{\m\n}\mc H^{0\m\n}
\non
&&-\frac{3}{2}\Sigma^{-2}\pd_\mu \Sigma \pd^\mu \Sigma +\frac{1}{16} \pd_\mu M_{MN}\pd^\mu M^{MN} -V +e^{-1}\mc L_\text{top}, \label{L-full}
\eeq
where $e$ is the vielbein determinant, and $\mc L_\text{top}$ is the topological term whose explicit form will not be given here due to its complexity and the fact that it is not relevant for the type of solutions considered in this work. The covariant gauge field strengths read
\begin{equation}
\mc{H}^{\mc{M}}_{\mu\nu}=2\pd_{[\mu}A^{\mc{M}}_{\nu]}+{X_{\mc{N}\mc{P}}}^{\mc{M}}A^{\mc{N}}_\mu A^{\mc{P}}_\nu\, .\label{HMmn}
\end{equation}
In the embedding tensor formalism, the covariant field strengths $\mc H^{\mc M}_{\m\n}$ originally contain some two-form fields $B_{\mu\nu \mc{M}}$ that are introduced off-shell, without any kinetic terms and coupling to vector fields only via a topological term. 
However, for the solutions considered here, the two-form fields can be consistently truncated out, in a manner similar to what was done in \cite{AdS5-bs}, \cite{AdS5-bh}.
\\\indent
The scalar potential in (\ref{L-full}) is given by
\begin{eqnarray}
V&=&-\frac{1}{4}\left[f_{MNP}f_{QRS}\Sigma^{-2}\left(\frac{1}{12}M^{MQ}M^{NR}M^{PS}-\frac{1}{4}M^{MQ}\eta^{NR}\eta^{PS}\right.\right.\nonumber \\
& &\left.+\frac{1}{6}\eta^{MQ}\eta^{NR}\eta^{PS}\right) +\frac{1}{4}\xi_{MN}\xi_{PQ}\Sigma^4(M^{MP}M^{NQ}-\eta^{MP}\eta^{NQ})\nonumber \\
& &\left.
+\frac{\sqrt{2}}{3}f_{MNP}\xi_{QR}\Sigma M^{MNPQRS}+\xi^M\xi^N\Sigma^{-2}M^{MN}\right] \label{V-o}
\end{eqnarray}
with $M^{MN}$ being the inverse of a symmetric matrix $M_{MN}$ defined by
\begin{equation}
M_{MN}=\mc{V}_M^{\phantom{M}m}\mc{V}_N^{\phantom{M}m}+\mc{V}_M^{\phantom{M}a}\mc{V}_N^{\phantom{M}a}\, .\label{Mmn}
\end{equation}
$M^{MNPQRS}$ is obtained from raising indices of $M_{MNPQR}$ defined by
\begin{equation}
M_{MNPQR}=\epsilon_{mnpqr}\mc{V}_{M}^{\phantom{M}m}\mc{V}_{N}^{\phantom{M}n}
\mc{V}_{P}^{\phantom{M}p}\mc{V}_{Q}^{\phantom{M}q}\mc{V}_{R}^{\phantom{M}r}\, .
\end{equation}
Equivalently, in terms of the gravitino $A^{ij}_1$, dilatino $A_2^{ij}$ and gaugino $A_{2}^{aij}$ shift matrices, 
\begin{eqnarray}
A_1^{ij}&=&\frac{1}{\sqrt{6}}\left(-\zeta^{(ij)} + 2\rho^{(ij)}\right), \label{A1f}\\
A_2^{ij}&=&\frac{1}{\sqrt{6}}\left(\zeta^{(ij)} + \rho^{(ij)}+\frac{3}{2}\tau^{[ij]} \right),\label{A2f} \\
A_2^{aij}&=&\frac{1}{2}\left(-\zeta^{a[ij]} + \rho^{a(ij)}-\frac{\sqrt{2}}{4}\tau^a\Omega^{ij}\right), \label{A2af}
\end{eqnarray}
where
\begin{eqnarray}
\tau^{[ij]}&=&\Sigma^{-1}{\mc{V}_M}^{ij}\xi^M,\non
\tau^a&=&\Sigma^{-1}{\mc{V}_M}^a\xi^M,
\non
\zeta^{(ij)}&=&\sqrt{2}\Sigma^2\Omega_{kl}{\mc{V}_M}^{ik}{\mc{V}_N}^{jl}\xi^{MN},\nonumber  \\ 
\zeta^{a[ij]}&=&\Sigma^2{\mc{V}_M}^a{\mc{V}_N}^{ij}\xi^{MN},
\non
\rho^{(ij)} &=&-\frac{2}{3}\Sigma^{-1}{\mc{V}^{ik}}_M{\mc{V}^{jl}}_N{\mc{V}^P}_{kl}{f^{MN}}_P, \nonumber \\
\rho^{a(ij)}&=&\sqrt{2}\Sigma^{-1}\Omega_{kl}{\mc{V}_M}^a{\mc{V}_N}^{ik}{\mc{V}_P}^{jl}f^{MNP},\,\label{Aq}
\end{eqnarray}
the scalar potential can be written as
\begin{equation}
V=-A_1^{ij}A_{1ij}+A_2^{ij}A_{2ij}+A_2^{aij}A^a_{2ij}\, .\label{V-a-sq}
\end{equation}
 Note also that lowering and raising of $USp(4)$ indices $i,j,\ldots$ with the symplectic form $\Omega_{ij}$ and its inverse $\Omega^{ij}$ correspond to complex conjugate for example $A_{1ij}=\Omega_{ik}\Omega_{jl}A_1^{kl}=(A_1^{ij})^*$. ${\mc{V}_M}^{ij}$ is related to ${\mc{V}_M}^m$ by $SO(5)$ gamma matrices\footnote{The explicit $SO(5)$ gamma matrices used are given in \cite{AdS5-bs}, \cite{AdS5-bh}. } and satisfies the relations
\begin{equation}
{\mc{V}_M}^{ij}=-{\mc{V}_M}^{ji}\qquad \textrm{and} \qquad {\mc{V}_M}^{ij}\Omega_{ij}=0.
\end{equation}
%%%%%%%%%%%%%%%%%%%%%%%%%%%%%%%%%%%%%%%%%%%%%%%%%%%%%%%%%%%
\section{A review of $dS_5$ solutions in 5D $N=4$ supergravity}\label{dS5-rev}
%%%%%%%%%%%%%%%%%%%%%%%%%%%%%%%%%%%%%%%%%%%%%%%%%%%%%%%%%%%
In  \cite{dS5}, using a novel method introducing two different sets of ansatze on the fermionic shift matrices\footnote{In (\ref{A2}, \ref{A2a}), $\langle\,\,\,\rangle$ denotes the evaluation of the enclosed quantity in the $dS_5$ vacuum.} (\ref{A1f}, \ref{A2f}, \ref{A2af}, \ref{Aq}) 
\beq
\langle A_1^{ij}\rangle =0, \qquad \langle A_2^{aij}\rangle =0, \qquad \langle A_2^{ik}A_{2jk}\rangle =\frac{1}{4}|\m|^2\d^i_j, \label{A2}\\
\langle A_1^{ij}\rangle =0, \qquad \langle A_2^{ij}\rangle =0, \qquad \langle A_2^{aik}A^a_{2jk}\rangle =\frac{1}{4}|\m|^2\d^i_j,
\label{A2a}
\eeq
that ensure the positivity of the scalar potential (\ref{V-a-sq}),
 the quadratic constraints (\ref{QC}) together with the extremum condition on the scalar potential were solved to obtain a large class of gaugings capable of admitting $dS_5$ solutions. This particular class corresponds to the ansatz (\ref{A2a}).  The ansatz (\ref{A2}), while satisfying the quadratic constraints (\ref{QC}), does not extremize the scalar potential, hence is eliminated from being eligible gaugings for $dS_5$. For more details on this method and the derivation of its associated solutions, see \cite{dS5}. In this section, we only recall the results obtained in \cite{dS5} that will be necessary for our ensuing discussion.
\\\indent
There are eight semisimple gaugings that can give rise to $dS_5$ solutions in $N=4$ supergravity. Non-semisimple gaugings were not considered in \cite{dS5} due to their inherently more complex nature compared to their semisimple counterparts. Each of the eight semisimple gaugings of \cite{dS5} is characterized by their embedding tensor components ($\xi_M, \xi_{MN}, f_{MNP}$) which are solutions of the quadratic constraints (\ref{QC}). The corresponding gauge generators are derived from the embedding tensor using (\ref{gauge-gen}). In all gauge groups, $\xi_M = 0$, and so the generators are given by
\beq
{X_{MN}}^P =-{f_{MN}}^P, \qquad {X_{0M}}^N = -{\xi_M}^N. \label{gauge-gen2}
\eeq
 All gaugings are non-compact and comprise two gauge factors of the form $SO(1,1)\times G_{nc}$, or $SO(1,1)^{(d)}_\text{diag}\times G_{nc}$. The compact part of $G_{nc}$ must be embedded in the matter symmetry group $SO(n)\subset SO(5,n)$ of the scalar manifold (\ref{sm}). This is in direct constrast to the case of those gaugings capable of admitting fully supersymmetric $AdS_5$ solutions in which the gauge group must comprise two factors of the form $U(1)\times G$ where $G$ must admit an $SU(2)$ subgroup that is itself a subgroup of the R-symmetry $SO(5)\subset SO(5,n)$ in (\ref{sm}).
 The gauge groups with $dS_5$ solutions and their corresponding embedding tensors are listed in the table below, which is originally presented in \cite{dS5}. All $dS_5$ solutions of \cite{dS5} are unstable due to the presence of tachyonic scalars in the spectrum. This is a common phenomenon for $dS$ vacua in gauged supergravity. 
\begin{table}[!htbp]
\centering
%\resizebox{\textwidth}{!}
\begin{tabular}{|c|l|c|}
\hline
 Gauge group $G$ & $\begin{array}{c}\text{Embedding tensor} \\ \,\,\xi_{MN},\,\,f_{MNP} \end{array}$ &  $\begin{array}{c}\text{Unbroken symmetry}\\ \text{of $dS_5$ vacuum}\end{array}$\\
\hline &&\\
$SO(1,1)\times SO(2,1)$ & $\xi_{56}=g_1,\quad f_{238} = -g_2$ &  $SO(2)$\\&&\\\hline&&\\
$SO(1,1)^{(2)}_\text{diag}\times SO(2,1)$ & $\begin{array}{l}\xi_{46}=\xi_{57} =g_1, \\f_{238} = -g_2\end{array}$ &  $SO(2)$\\&&\\\hline&&\\
$SO(1,1)^{(3)}_\text{diag}\times SO(2,1)$ & $\begin{array}{l}\xi_{19}=\xi_{56}=\xi_{47} =g_1, \\f_{238} = -g_2\end{array}$ & $SO(2)$\\&&\\\hline&&\\
 $SO(1,1)\times SO(2,2)$ & $\begin{array}{l}\xi_{56}=g_1, \\f_{238} =-g_2,\, f_{149} = -g_3\end{array}$  & $SO(2)\times SO(2)$\\&&\\\hline&&\\
$SO(1,1)\times SO(3,1)$ &$\begin{array}{l}\xi_{56}=g_1, \\f_{8,9,10}=-g_2, \\f_{128} = f_{139} = f_{2,3,10}=g_2\end{array}$ &$SO(3)$ \\&&\\\hline&&\\
$SO(1,1)^{(2)}_\text{diag}\times SO(3,1)$  &$\begin{array}{l}\xi_{46}=\xi_{57}=g_1, \\f_{8,9,10}=-g_2, \\f_{128} =f_{139} = g_2,\\ f_{2,3,10}=g_2\end{array}$ & $SO(3)$\\&&\\\hline &&\\
 $SO(1,1)\times SU(2,1)$ & $\begin{array}{l}\xi_{56} =1,\\f_{129}=f_{138} =g_2,\\ f_{147} = f_{248} = g_2,\\ f_{349} = f_{237} = -g_2,\\ f_{789} = -2g_2, \\f_{3,4,10} = f_{1,2,10} = \sqrt{3}g_2\end{array}$  & $SU(2)\times U(1)$ \\&&\\\hline&&\\
$SO(1,1)\times SO(4,1)$ & $\begin{array}{l}\xi_{56} =1, \\f_{127} = f_{138} = f_{1,4,10} = g_2, \\ f_{239} =f_{2,4,11} = f_{3,4,12} = g_2,\\ f_{789} = f_{7,10,11}=-g_2,\\  f_{8,10,12} = f_{9,11,12} = -g_2,\end{array}$  &$SO(4)$ \\&&\\
\hline
\end{tabular}
\caption{The eight gauge groups that give rise to $dS_5$ vacua in matter-coupled $N=4$ five-dimensional gauged supergravity and their corresponding embedding tensors. This table is originally presented in \cite{dS5}.}\label{table:dS5}
\end{table}
\newpage
For later convenience, we also provide a summary of all scalar potentials with $dS_5$ solution arising from the eight gauge groups listed in Table \ref{table:dS5}, together with the $g_1/g_2$ scaling ratios needed to bring the $dS_5$ vacua to the origin of the scalar manifold (\ref{sm}). Note that $g_1$ and $g_2$ are the gauge constants of the two gauge factors in the gauge group $G$. Given the explicit forms of the embedding tensors $f_{MNP}, \xi_{MN}$, the scalar potential can be constructed using (\ref{V-o}) or (\ref{V-a-sq}). 
\\\indent
All $dS_5$ vacua resulting from these gauge groups can be derived with all scalars from the vector multiplets truncated out. Consequently,  the scalar potential $V$ is only a function of the dilaton $\S$. 
We also note that  in each gauge group there are two equivalent choices for the $g_1/g_2$ ratio, corresponding to either a plus or a minus sign. This results in two equivalent gauged theories with the same gauge group and the same $dS_5$ vacuum. Without loss of generality, we can choose the $+$ sign from here on. 
In subsequent sections, we will make use of the embedding tensors and scalar potentials listed in the tables (\ref{table:dS5}, \ref{table:dS5-V}) to construct the eight different gauged theories and derive their corresponding $dS_{5-d}\times \S^d$ ($d=2,3$) solutions. 
%%%%%%%%%%%%%%%%%%%%%%%%%%%%%%%%%%%%%%%%%%%%%%%%%%%%%%%%%%%%%%
\begin{table}[!htbp]
\centering
%\resizebox{\textwidth}{!}
\begin{tabular}{|c|c|c|}
\hline
Gauge group $G$& Scalar potential $V$ & $g_1/g_2$ ratio
\\\hline &&\\
$SO(1,1)\times SO(2,1)$ & $\dfrac{2g_2^2 + g_1^2 \Sigma^6}{4\Sigma^2}$  & $g_1 = \pm g_2$\\&&\\\hline &&\\
$SO(1,1)\times SO(2,2)$ & $\dfrac{4g_2^2 + g_1^2\S^6}{4\S^2}$ & $g_1 = \pm \sqrt{2}g_2$\\&&\\\hline &&\\
$SO(1,1)\times SO(3,1$ & $\dfrac{6g_2^2 + g_1^2\S^6}{4\S^2}$ & $g_1 = \pm \sqrt{3}g_2$\\&&\\\hline &&\\
$SO(1,1)\times SO(4,1)$ & $\dfrac{12g_2^2 + g_1^2\S^6}{4\S^2}$ & $g_1 = \pm \sqrt{6}g_2$\\&&\\\hline &&\\
$SO(1,1)\times SU(2,1)$ & $\dfrac{24g_2^2 + g_1^2\S^6}{4\S^2}$ & $g_1 = \pm 2\sqrt{3}g_2 $\\&&\\
\hline &&\\
$SO(1,1)^{(2)}_\text{diag}\times SO(2,1)$ & $\dfrac{g_2^2 + g_1^2\S^6}{2\S^2}$ & $g_1 = \pm \dfrac{1}{\sqrt{2}}g_2$\\&&\\\hline &&\\
$SO(1,1)^{(3)}_\text{diag}\times SO(2,1)$ & $\dfrac{2g_2^2 + 3g_1^2\S^6}{4\S^2}$ & $g_1 = \pm \dfrac{1}{\sqrt{3}}g_2$\\&&\\\hline &&\\
$SO(1,1)^{(2)}_\text{diag}\times SO(3,1)$ & $\dfrac{3g_2^2 + g_1^2\S^6}{2\S^2}$ & $g_1 = \pm \dfrac{\sqrt{3}}{\sqrt{2}}g_2$\\&&
\\\hline
\end{tabular}
\caption{Scalar potentials of the eight gauge groups with $dS_5$ solutions. Each potential can admit a $dS_5$ solution at the origin of the scalar manifold $\S=1$ with the indicated $g_1/g_2$ scaling.}\label{table:dS5-V}
\end{table}
\newpage

%%%%%%%%%%%%%%%%%%%%%%%%%%%%%%%%%%%%%%%%%%%%%%%%%%%%%%%%%%%
\section{Cosmological solutions from 5D $N=4$ supergravity}\label{ans}
%%%%%%%%%%%%%%%%%%%%%%%%%%%%%%%%%%%%%%%%%%%%%%%%%%%%%%%%%%%
In this section we specify our ansatze for the various fields in the 5D $N=4$ theory that are required to derive cosmological solutions. For this type of solutions, we only turn on the metric $g_{\m\n}$, the dilaton $\S$, and some relevant gauge fields $A^M_\m$ while all other fields are kept vanishing. Specifically,
\beq
A^0 = 0, \qquad \phi^{m a} = 0.
\eeq
The vanishing of the gauge field $A^0$ means that $\mc H^0 = 0$, while
the vanishing of the vector multiplet scalars $\p^{m a}$ means that the symmetric matrix $M_{MN}$ (\ref{Mmn}) and its inverse $M^{MN}$ reduce to just the $(5+n)\times (5+n)$ identity matrices.
\beq
M_{MN} = M^{MN} = \mf 1_{5+n}.
\eeq
The  bosonic Lagrangian (\ref{L-full}) reduces to
\beq
e^{-1}\mc{L}&=&\frac{1}{2}R -\frac{1}{4}\S^2\delta_{MN}\mc H_{\m\n}^M \mc H^{N\m\n} 
-\frac{3}{2}\Sigma^{-2}\pd_\mu \Sigma \pd^\mu \Sigma  -V.\label{L-red}
\eeq
%%%%%%%%%%%%%%%%%%%%%%%%%%%%%%%%%%%%%%%%%%%%%%%%%%%%%%%%%%%
\subsection{$dS_3\times \Sigma_2$ solutions}
%%%%%%%%%%%%%%%%%%%%%%%%%%%%%%%%%%%%%%%%%%%%%%%%%%%%%%%%%%%
We begin with the metric ansatz
\beq
ds^2 = -dt^2 + e^{2f}(dx^2 + dy^2) + e^{2 g}d\Omega_2^2 \label{dS3-m-ans}
\eeq
where $f, g$ are functions of the time variable $t$, and $d\Omega^2_2$ is the line element of $\Sigma_2 = S^2, H^2$.
\beq
d\O_2^2 = \begin{dcases} d\theta^2 + \sin^2\theta \,d\phi^2, & S^2\\  d\theta^2 + \sinh^2\theta \,d\phi^2, &H^2\end{dcases}\,\,. \label{O2}
\eeq
 We will turn on an Abelian gauge field $U(1)\cong SO(2)\subset G_{nc} \subset SO(n)$, where $SO(n)$ is the matter symmetry group in (\ref{sm}), and $G_{nc} =SO(2,1), SO(2,2), SO(3,1), SO(4,1), SU(2,1)$, with the ansatz
\beq
A^M = \begin{dcases}a \cos \theta\, d\phi, & S^2\\ a\cosh\theta d\phi, & H^2 \end{dcases}\,, \label{dS3-A-ans}
\eeq
where the index $M$ is specific to each gauged theory and will be specified accordingly. 
The field strength is given by (\ref{HMmn}) and is simply
\beq
\mc H^M_{\theta\phi} = \begin{dcases} a\sin\theta, & S^2 \\ a\,\sinh\theta, & H^2\end{dcases} \,\,. \label{HdS3}
\eeq
Substituting the ansatze (\ref{dS3-m-ans}) and (\ref{HdS3}) in the Lagrangian (\ref{L-red}) results in the following second-order equations of motion
\beq
0 &=&\lambda e^{-2 g}-\frac{1}{2} a^2 e^{-4 g} \S^2+\ddot f+2 \dot f \dot g+\dot f^2+2 \ddot g+3 \dot g^2 +\frac{3 \dot \S^2}{2 \S^2}-V(\S) \non
0&=& \frac{1}{2} a^2 e^{-4 g} \S^2+2 \ddot f+2 \dot f \dot g+3 \dot f^2+\ddot g+\dot g^2+\frac{3 \dot \S^2}{2 \S^2}-V(\S)\non
0&=& -\frac{1}{2} a^2 e^{-4 g} \S^2 - \dot \S \left(\frac{3 \dot f}{\S}+\frac{3 \dot g}{\S}\right)-\frac{3 \ddot \S}{2 \S}+\frac{3 \dot \S^2}{2 \S^2}-\frac{1}{2} \S \,V'(\S) \label{dS3eom}
\eeq
where 
\beq
\lambda = \begin{dcases} +1, & \S_2 = S^2 \\ -1, & \S_2 = H^2 \end{dcases}\,\,.
\eeq
A $dS_3\times \Sigma_2$ solution of this theory arises as the fixed point solution
\beq
f(t) = f_0 \,t, \qquad g(t) = g_0, \qquad \Sigma(t) = \Sigma_0\,\,. \label{dS3fp}
\eeq
The full solution described by (\ref{dS3-m-ans}) is a cosmological solution that interpolates between the above $dS_3\times \S_2$ fixed point (\ref{dS3fp}) at $t\ra -\infty$ and the unique $dS_5$ solution of each theory at $t\ra \infty$. In all of the gauged theories studied in this work, we find that in order to obtain real solutions, the following condition needs to be imposed on the gauge flux $a$ and gauge constant $g_2$
\beq
a^2 g^2_2 = -1, \qquad \text{or} \qquad a\, g_2 = \pm i. \label{dS3twist}
\eeq
This resembles the situation in \cite{Lu-03} for cosmological solutions of the $dS$ supergravities, with the wrong signs for the kinetic terms of the gauge field strengths, that arise from the dimensional reduction of the exotic  M$\star$ and IIB$\star$ theories. Given that our theory under consideration is a conventional supergravity with the correct signs for the gauge field strengths, this feature is rather puzzling. In section \ref{concl}, we will comment further on this. 
%%%%%%%%%%%%%%%%%%%%%%%%%%%%%%%%%%%%%%%%%%%%%%%%%%%%%%%%%%%
\subsection{$dS_2\times \Sigma_3$ solutions}
%%%%%%%%%%%%%%%%%%%%%%%%%%%%%%%%%%%%%%%%%%%%%%%%%%%%%%%%%%%
The ansatz for the metric is
\beq
ds^2 = -dt^2 + e^{2f} dr^2 + e^{2g} d\Omega_3^2,
\label{dS2-m-ans} 
\eeq
where $f, g$ are functions of the time variable $t$, and $d\Omega_3^2$ is the line element for $S^3$ and $H^3$. 
\beq
d\O_3^2 = \begin{dcases} d\psi^2 + \sin^2\psi\lf( d\theta^2 + \sin^2\theta d\phi^2\rr) & (S^3)\\ d\psi^2 + \sinh^2\psi\lf( d\theta^2 + \sin^2\theta d\phi^2\rr)& (H^3)\end{dcases}\,\,. \label{O3}
\eeq
For the gauge fields, we will need to turn on those corresponding to a full non-Abelian $SO(3)$ factor. In this case, unlike the $dS_3$ solution that is present in all $dS_5$ gauged theories, only four theories, $SO(1,1)\times SO(3,1)$, $SO(1,1)^{(2)}_\text{diag}\times SO(3,1)$, $SO(1,1)\times SO(4,1)$ and  $SO(1,1)\times SU(2,1)$ can admit this solution since their second gauge factor  $G_{nc}$ is large enough to admit the $SO(3)$ factor as a subgroup. The general ansatz for the gauge fields is
\beq
\begin{dcases} A^M_\theta  = a_M \cos\psi, \,\,\, A^N_\phi = a_N \cos\theta,\,\,\, A^P_\phi = a_P\cos \psi \sin\theta, & \S_3 = S^3 
\\
A^M_\theta = a_M \cosh\psi, \,\,\, A^N_\phi = a_N \cos\theta, \,\,\, A^{P}_\phi = a_{P} \cosh\psi\, \sin\theta, & \S_3 = H^3 
 \end{dcases}
\label{dS2-A-ans}
\eeq
where $A^M, A^N, A^P$ are the gauge fields correspond to the generators $X^M, X^N, X^P$ (\ref{gauge-gen2}) of an $SO(3)$ subgroup of the gauge factors $SO(3,1), SO(4,1)$ or $SU(2,1)$. 
The covariant gauge field strengths are given by (\ref{HMmn}) but with a factor of $\pm i$ in front of the structure constant ${X_{NP}}^P$.
\beq
\mc H^M_{\m\n} = 2 \pd_{[\m} A^M_{\n]} \pm i\, {X_{ N  P}}^{ M} A^{N}_\m A^P_\n. \label{HMmni}
\eeq
We note that the factor $\pm i$ in (\ref{HMmni}) does not amount to an analytic continuation $g_k \ra i\, g_k$, where $k=1,2$, or a modification of the gauged theory in any way. Rather, it serves as a means to impose the  relation (\ref{dS2twist}) below between the gauge flux and the gauge constant $g_2$, analogous to (\ref{dS3twist}), in order to obtain real solutions. 
\\\indent
After setting $a_M = a_N =a_P =a$, and imposing the condition\footnote{ to make the field strength component $\mc H^P_{\theta\phi}$ vanish.}
\beq
a = \mp \frac{i}{c\,g_2}\,. \label{dS2twist}
\eeq
where $c$ is a constant specific to each gauging, in (\ref{HMmni}), the  resulting covariant field strengths corresponding to (\ref{dS2-A-ans}) are given by 
\beq
\begin{dcases} 
\mc H^M_{\psi\theta} = a \sin\theta, \,\,\,\mc H^N_{\theta\phi} = a \sin\theta\sin^2\psi, \,\,\, \mc H^P_{\psi\phi} = a\sin\theta \sin\psi, & \S_3 = S^3\\
\mc H^M_{\psi\theta} =a \sinh\psi, \,\,\,
\mc H^N_{\theta\phi} = a \sin\theta \sinh\psi^2, \,\,\,
\mc H^{P}_{\psi\phi}= a \sin\theta \sinh\psi 
, & \S_3 = H^3
\end{dcases}\,\,.
\label{HdS2}
\eeq
 The $\mp$ signs in \ref{dS2twist} are correlated with the $\pm$ sign in (\ref{HMmni}). The condition (\ref{dS2twist}) are needed to guarantee the reality of our $dS_2\times \S_3$ solutions. Without loss of generality, we can choose the  $-$ sign in (\ref{HMmni}) such that (\ref{dS2twist}) becomes
 \beq
a = \frac{i}{ c\,g_2}.  \label{dS2twist2}
\eeq
Substituting the ansatze (\ref{dS2-m-ans}) and (\ref{HdS2}) in (\ref{L-red}) results in the following equations of motion
\beq
0 &=& \lambda e^{-2 g}-\frac{1}{2}a^2 e^{-4 g} \S^2+\ddot g+2 \dot g^2 +\frac{\dot \S^2}{2 \S^2}-\frac{1}{3} V(\S)\non
0&=&  \lambda e^{-2 g}+\frac{1}{2}a^2 e^{-4 g} \S^2+\ddot f+2 \dot f \dot g+\dot f^2+2 \ddot g+3 \dot g^2 +\frac{3 \dot \S^2}{2 \S^2}-V(\S)
\non
0&=& - a^2 e^{-4 g} \S^2-\frac{\dot \S \left(\dot f+3 \dot g\right)}{\S}-\frac{\ddot \S}{\S}+\frac{\dot \S^2}{\S^2}-\frac{1}{3} \S\, V'(\S), \label{dS2eom}
\eeq
with 
\beq
\lambda = \begin{dcases} +1, & \S_3 = S^3\\ -1, & \S_3 = H^3 \end{dcases}\,\,.
\eeq
The $dS_2\times \Sigma_3$ solution of this theory arises as the fixed point solution
\beq
f(t) = f_0 \,t, \qquad g(t) = g_0, \qquad \Sigma(t) = \Sigma_0. \label{dS2fp}
\eeq
The full solution described by \ref{dS2-m-ans} is a cosmological solution that interpolates between the above $dS_2\times \S_3$ fixed point (\ref{dS2fp}) at $t\ra -\infty$ and the unique $dS_5$ solution of each theory at $t\ra \infty$. 
%%%%%%%%%%%%%%%%%%%%%%%%%%%%%%%%%%%%%%%%%%%%%%%%%%%%%%%%%%%
\section{All $dS$ gauged theories} \label{all-G}
%%%%%%%%%%%%%%%%%%%%%%%%%%%%%%%%%%%%%%%%%%%%%%%%%%%%%%%%%%%
In the following subsections, the $dS_5$ solution of each gauged theory 
\beq
ds^2 &=& -dt^2 + e^{2f_0 t}(dx^i)^2, \qquad i = 1, \ldots, 4\non
\S(t)  &=& \S_0.
\eeq
will be specified by $(\S_0, f_0)$ 
where $f_0$ is related to the extremized value $V_0$ of the scalar potential as
\beq
f_0 = \sqrt{\frac{V_0}{6}}.
\eeq
 %%%%%%%%%%%%%%%%%%%%%%%%%%%%%%%%%%%%%%%%%%%%%%%%%%%%%%%%%%%
\subsection{$SO(1,1)\times SO(2,1)$}
%%%%%%%%%%%%%%%%%%%%%%%%%%%%%%%%%%%%%%%%%%%%%%%%%%%%%%%%%%%
The scalar potential of this gauged theory is
\beq
V = \frac{2 g_2^2 + g_1^2\S^6}{4\S^2} \label{Vso21}
\eeq
with the following $dS_5$ vacuum
\beq
\S_0 = 1, \qquad f_0 = \frac{1}{2\sqrt{2}}g_2, \qquad g_1= g_2.\label{so21-dS5}
\eeq
The embedding tensor for this theory is given in Table \ref{table:dS5}.
%%%%%%%%%%%%%%%%%%%%%%%%%%%%%%%%%%%%%%%%%%%%%%%%%%%%%%%%%%%
\subsubsection{$dS_3\times \Sigma_2$}
%%%%%%%%%%%%%%%%%%%%%%%%%%%%%%%%%%%%%%%%%%%%%%%%%%%%%%%%%%%
To obtain $dS_3\times \S_2$ solutions, we turn on the gauge field $A^8$ corresponding to the gauge generator $X_8$ of the $SO(2)\subset SO(2,1)$ factor (see Table \ref{table:dS5}). The gauge ansatz is given by (\ref{dS3-A-ans}) with $M=8$. 
The equations (\ref{dS3eom}) together with the scalar potential (\ref{Vso21}) have the following $dS_3\times H^2$ fixed point solution
\beq
\begin{array}{l}\S_0= 2^{1/6} \,\zeta,\\\\
f_0= 2^{-2/3}\, g_2 \,\zeta^2,\\\\
g_0= \dfrac{1}{2} \log \left\{2^{1/3} \left(\sqrt{1-a^2 g_2^2}-1\right)\,\dfrac{\zeta^2}{g_2^2}\right\},\\\\
 \zeta=\lf[{\dfrac{a^2 g_2^2-\sqrt{1-a^2 g_2^2}-1}{a^2 g_2^2}}\rr]^{1/6},
 \end{array} \label{so21-dS3-0}
\eeq
which becomes real,
\beq
\S_0&=& \sqrt[6]{2 \left(\sqrt{2}+2\right)},\non
f_0&=& \frac{\sqrt[3]{\sqrt{2}+2} }{2^{2/3}}\,g_2,\non
g_0&=& \frac{1}{2} \log \left(\frac{\left(\sqrt{2}-1\right) \sqrt[3]{2 \left(\sqrt{2}+2\right)}}{g_2^2}\right) \label{so21-dS3}
\eeq
after imposing (\ref{dS3twist}).
%%%%%%%%%%%%%%%%%%%%%%%%%%%%%%%%%%%%%%%%%%%%%%%%%%%%%%%%%%%
\subsubsection{Cosmological solution between $dS_3\times \Sigma_2$ and $dS_5$}
\begin{figure}[!htb]
\centering 
  \begin{subfigure}[b]{0.3\textwidth}
    \includegraphics[width=\textwidth]{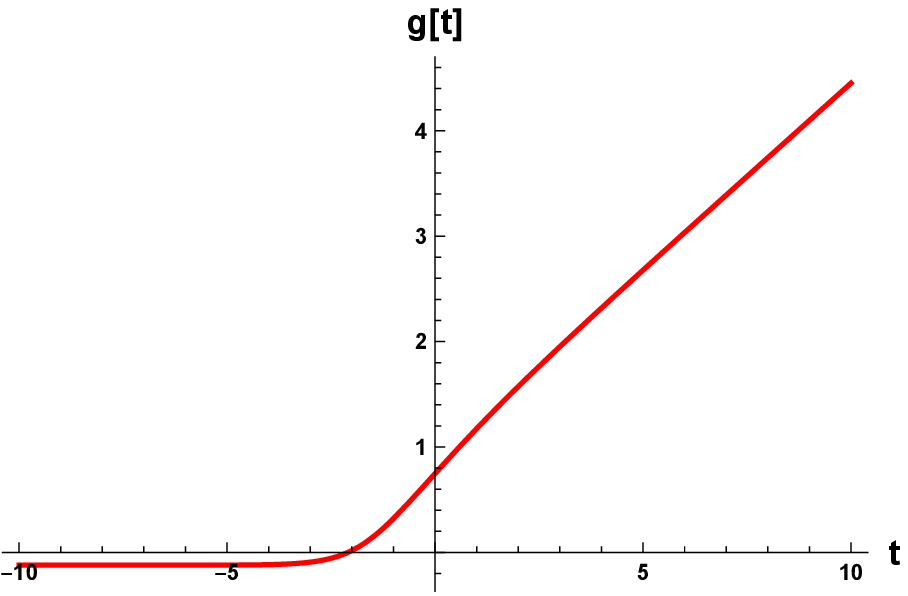}
\caption{Solution for $g$}  
  \end{subfigure}
      \begin{subfigure}[b]{0.3\textwidth}
    \includegraphics[width=\textwidth]{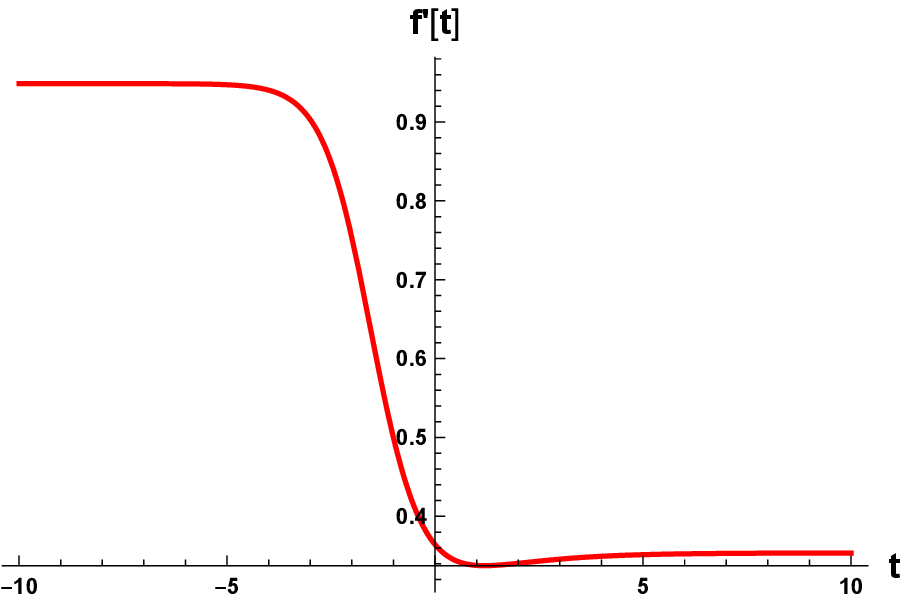}
\caption{Solution for $\dot f$}
    \end{subfigure}
      \begin{subfigure}[b]{0.3\textwidth}
    \includegraphics[width=\textwidth]{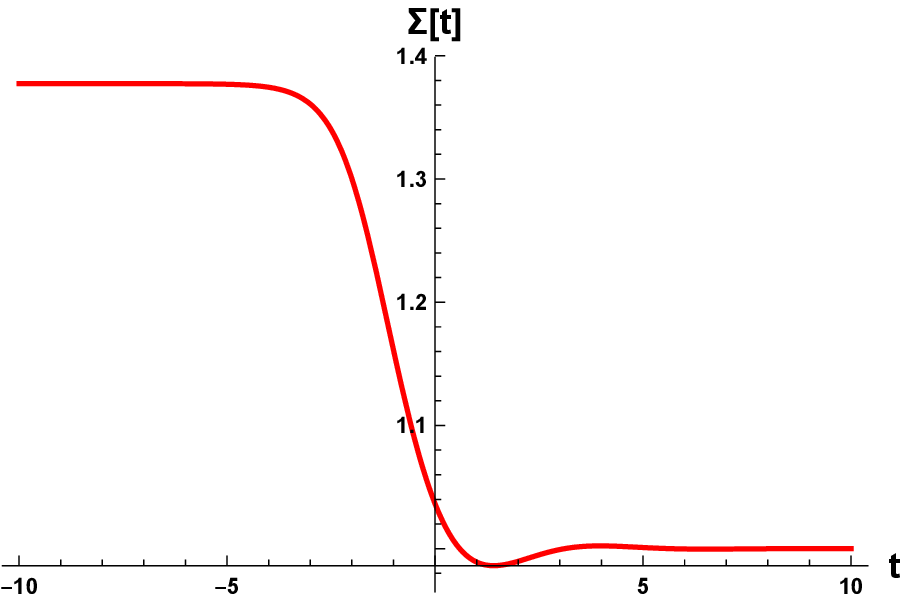}
\caption{Solution for $\Sigma$}
  \end{subfigure}
    \caption{A cosmological  solution from $dS_3\times H^2$ (\ref{so21-dS3}) at the infinite past to the $dS_5$ solution (\ref{so21-dS5}) in the infinite future from $SO(1,1)\times SO(2,1)$ gauge group and $g_2 = 1$.}
    \label{fig:SO21-dS3-cf}
    \end{figure}

%%%%%%%%%%%%%%%%%%%%%%%%%%%%%%%%%%%%%%%%%%%%%%%%%%%%%%%%%%%
Numerically solving the equations (\ref{dS3eom}) with the potential (\ref{Vso21}) gives us the following interpolating solution plotted in Fig.\ref{fig:SO21-dS3-cf}. 
%\newpage

%%%%%%%%%%%%%%%%%%%%%%%%%%%%%%%%%%%%%%%%%%%%%%%%%%%%%%%%%%%
\subsection{$SO(1,1)^{(n)}_\text{diag}\times SO(2,1)$ with $n=2,3$}
%%%%%%%%%%%%%%%%%%%%%%%%%%%%%%%%%%%%%%%%%%%%%%%%%%%%%%%%%%%
These theories have identical $dS_5$ solutions to (\ref{so21-dS5}) of the $SO(1,1)\times SO(2,1)$ gauged theory. The only difference among these three theories is the ratio $g_1/g_2$ which scales the $dS_5$ solution to the origin of the scalar manifold. 
\ben
\item $SO(1,1)^{(2)}_\text{diag}\times SO(2,1)$: The scalar potential is
\beq
V = \frac{ g_2 + \, g_1 \S^6}{2\S^2}
\eeq
with the following $dS_5$ point
\beq
\S_0 = 1, \qquad f_0 =  \frac{1}{2\sqrt{2}g_2}\qquad g_1 =  \frac{1}{\sqrt{2}}\,{g_2}.\eeq

\item $SO(1,1)^{(3)}_\text{diag}\times SO(2,1)$: The scalar potential is
\beq
V = \frac{2 g_2 + 3 g_1 \S^6}{4\S^2}
\eeq
with the following $dS_5$ point
\beq
\S_0 = 1, \qquad f_0 = \frac{1}{2\sqrt{2}g_2}, \qquad g_1 =  \frac{1}{\sqrt{3}}{g_2}.
\eeq
\enn
The embedding tensors for the $SO(2,1)$ part of these theories are identical to that of the $SO(1,1)\times SO(2,1)$ theory, see Table \ref{table:dS5}. 
To obtain $dS_3\times \S_2$ solution, we again turn on the gauge field $A^8$ corresponding to the gauge generator $X_8$ of the $SO(2)\subset SO(2,1)$ factor. The gauge ansatz is given by (\ref{dS3-A-ans}) with $M=8$. 
The $dS_3\times H^2$ solutions of these theories are given by (\ref{so21-dS3-0}, \ref{so21-dS3}) and the cosmological solution is again given in Fig. \ref{fig:SO21-dS3-cf}.
%%%%%%%%%%%%%%%%%%%%%%%%%%%%%%%%%%%%%%%%%%%%%%%%%%%%%%%%%%%
\subsection{$SO(1,1)\times SO(2,2)$}
%%%%%%%%%%%%%%%%%%%%%%%%%%%%%%%%%%%%%%%%%%%%%%%%%%%%%%%%%%%
The scalar potential of this gauged theory is
\beq
V=\frac{4g_2 + g_1 \S^6}{4\S^2} \label{Vso22}
\eeq
with the following $dS_5$ vacuum
\beq
\S_0 = 1, \qquad f_0 = \frac{g_2}{2}, \qquad g_1 =  \sqrt{2}g_2.\label{so22-dS5}
\eeq
The embedding tensor for this theory is given in Table \ref{table:dS5}.
%%%%%%%%%%%%%%%%%%%%%%%%%%%%%%%%%%%%%%%%%%%%%%%%%%%%%%%%%%%
\subsubsection{$dS_3\times \Sigma_2$}
%%%%%%%%%%%%%%%%%%%%%%%%%%%%%%%%%%%%%%%%%%%%%%%%%%%%%%%%%%%
For this gauged theory, since $SO(2,2)\cong SO(2,1)_1\times SO(2,1)_2$, one can turn on either $A^{M=8}$ or $A^{M=9}$ in (\ref{dS3-A-ans}) corresponding to $X_8$ or $X_9$, which generate the $SO(2)\subset SO(2,1)_1$ or $SO(2)\subset SO(2,1)_2$, respectively. Either choice leads to the same equations of motion (\ref{dS3eom}), which, together with the scalar potential (\ref{Vso22}) yield the following $dS_3\times H^2$ fixed point 
\beq
\begin{array}{l}
\S_0= \zeta,\\
f_0= \dfrac{1}{\sqrt{2}}\,g_2 \,\zeta^2,\\
g_0= \dfrac{1}{2} \log \left\{\dfrac{1}{2}\left(\sqrt{1-2 a^2 g_2^2}-1\right) \, \dfrac{\zeta^2}{g_2^2}\right\},\\
\zeta = \lf[\dfrac{2 a^2 g_2^2-\sqrt{1-2 a^2 g_2^2}-1}{a^2 g_2^2}\rr]^{1/6},
\end{array} \label{so22-dS3-0}
\eeq
which is real and becomes
\beq
\S_0&=& \sqrt[6]{\sqrt{3}+3},\non
f_0&=& \frac{\sqrt[3]{\sqrt{3}+3}}{\sqrt{2}}\, g_2,
\non
g_0&=& \frac{1}{2} \log \left(\frac{\left(\sqrt{3}-1\right) \sqrt[3]{\sqrt{3}+3}}{2 g_2^2}\right)\label{so22-dS3}
\eeq
upon imposing (\ref{dS3twist}).
%%%%%%%%%%%%%%%%%%%%%%%%%%%%%%%%%%%%%%%%%%%%%%%%%%%%%%%%%%%
\subsubsection{Cosmological solution between $dS_3\times \Sigma_2$ and $dS_5$}
\begin{figure}[!htb]
\centering 
  \begin{subfigure}[b]{0.3\textwidth}
    \includegraphics[width=\textwidth]{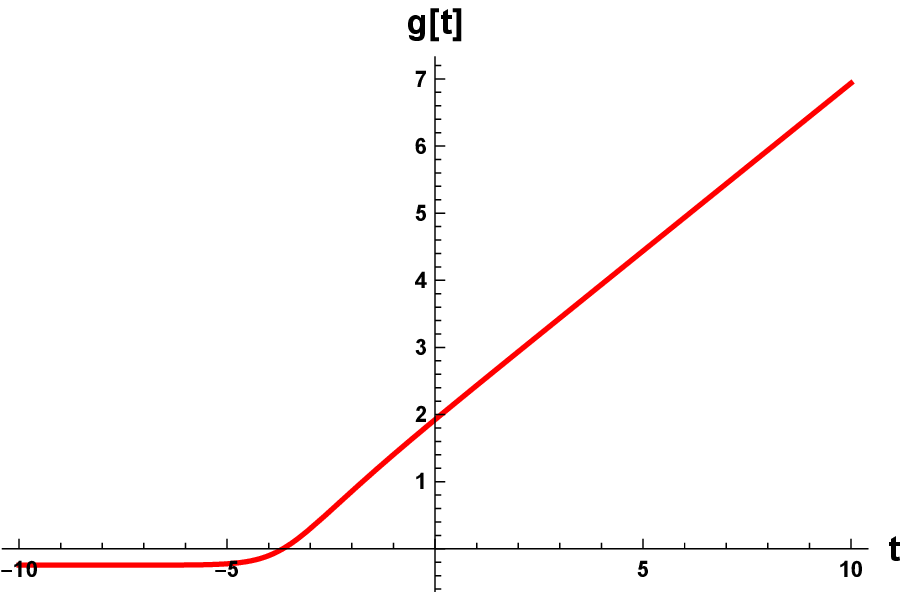}
\caption{Solution for $g$}  
  \end{subfigure}
      \begin{subfigure}[b]{0.3\textwidth}
    \includegraphics[width=\textwidth]{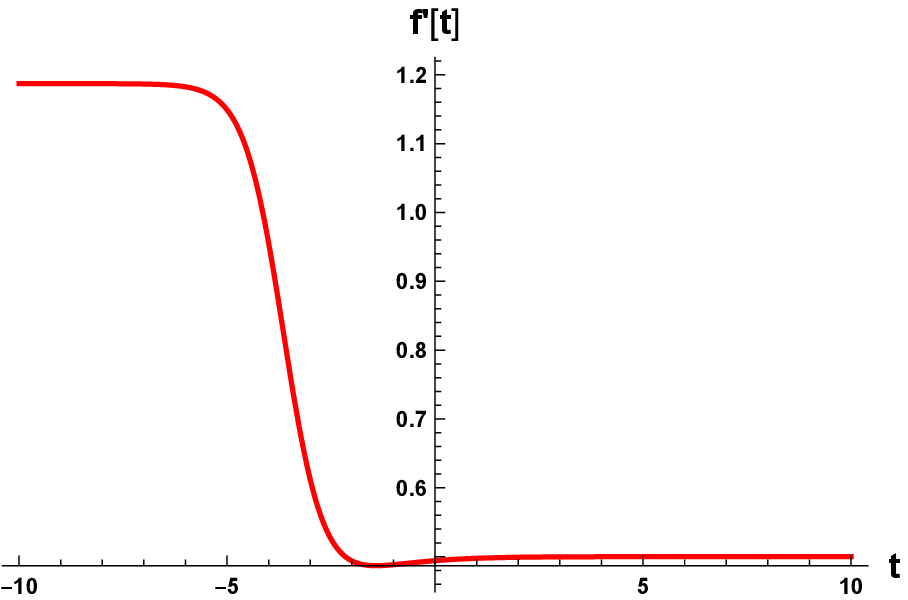}
\caption{Solution for $\dot f$}
    \end{subfigure}
    \begin{subfigure}[b]{0.3\textwidth}
    \includegraphics[width=\textwidth]{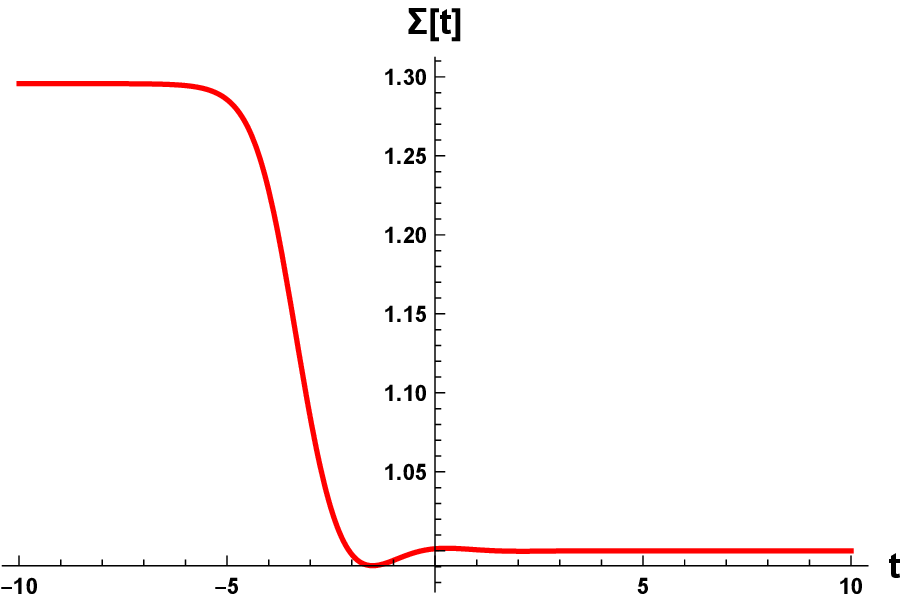}
\caption{Solution for $\Sigma$}
  \end{subfigure}
    \caption{A cosmological  solution from $dS_3\times H^2$ (\ref{so22-dS3}) at the infinite past to the $dS_5$ solution (\ref{so22-dS5}) in the infinite future from $SO(1,1)\times SO(2,2)$ gauge group and $g_2 = 1$.}
    \label{fig:SO22-dS3-cf}
    \end{figure}

%%%%%%%%%%%%%%%%%%%%%%%%%%%%%%%%%%%%%%%%%%%%%%%%%%%%%%%%%%%
We now numerically solve (\ref{dS3eom}) with the potential (\ref{Vso22}) to obtain the interpolating solution plotted in Fig.\ref{fig:SO22-dS3-cf}. 
    \newpage
%%%%%%%%%%%%%%%%%%%%%%%%%%%%%%%%%%%%%%%%%%%%%%%%%%%%%%%%%%%
\subsection{$SO(1,1)\times SO(3,1)$}
%%%%%%%%%%%%%%%%%%%%%%%%%%%%%%%%%%%%%%%%%%%%%%%%%%%%%%%%%%%
This gauged theory is characterized by the following scalar potential
\beq
V = \frac{6g_2^2 + g_1^2 \S^6}{4\S^2} \label{Vso31}
\eeq
with the following $dS_5$ critical point
\beq
\S_0 = 1, \qquad f_0 = \frac{1}{2}\sqrt{\frac{3}{2}}g_2, \qquad g_1 =  \sqrt{3}g_2.\label{so31-dS5}
\eeq
The embedding tensor for this theory is given in Table \ref{table:dS5}.
%%%%%%%%%%%%%%%%%%%%%%%%%%%%%%%%%%%%%%%%%%%%%%%%%%%%%%%%%%%
\subsubsection{$dS_3\times \Sigma_2$}
%%%%%%%%%%%%%%%%%%%%%%%%%%%%%%%%%%%%%%%%%%%%%%%%%%%%%%%%%%%
For this type of solutions, we turn on the $U(1)$ gauge field $A^{M=10}$ (\ref{dS3-A-ans}) corresponding to $X^{10}$ of the $U(1)\subset SO(3)\subset SO(3,1)$ factor. 
The equations (\ref{dS3eom}) with the potential (\ref{Vso31}) produce the following $dS_3\times H^2$ fixed point
\beq
\begin{array}{l}
\S_0= \lf(\dfrac{2}{3}\rr)^{1/6} \,\zeta, \\
f_0= \lf(\dfrac{\sqrt{3}}{4}\rr)^{1/3} \,g_2 \,\zeta^2,
\\
g_0= \dfrac{1}{2} \log \left(\dfrac{2^{1/3} }{3^{4/3} } \left(\sqrt{1-3 a^2 g_2^2}-1\right)\,\dfrac{\zeta^2}{g_2^2}\right), 
\\
\zeta = \lf[\dfrac{3 a^2 g_2^2-\sqrt{1-3 a^2 g_2^2}-1}{a^2 g_2^2}\rr]^{1/6}\end{array}, \label{so31-dS3-0}
\eeq
which is real after imposing (\ref{dS3twist})
\beq
\S_0&=& \sqrt[3]{2}, \non
f_0&=& \frac{\sqrt{3} }{\sqrt[3]{2}}\,g_2,\non
g_0&=& \frac{1}{2} \log \left(\frac{2^{2/3}}{3 g_2^2}\right)\,\,. \label{so31-dS3}
\eeq
%%%%%%%%%%%%%%%%%%%%%%%%%%%%%%%%%%%%%%%%%%%%%%%%%%%%%%%%%%%
\subsubsection{Cosmological solution between $dS_3\times \Sigma_2$ and $dS_5$}
\begin{figure}[!htb]
\centering 
  \begin{subfigure}[b]{0.3\textwidth}
    \includegraphics[width=\textwidth]{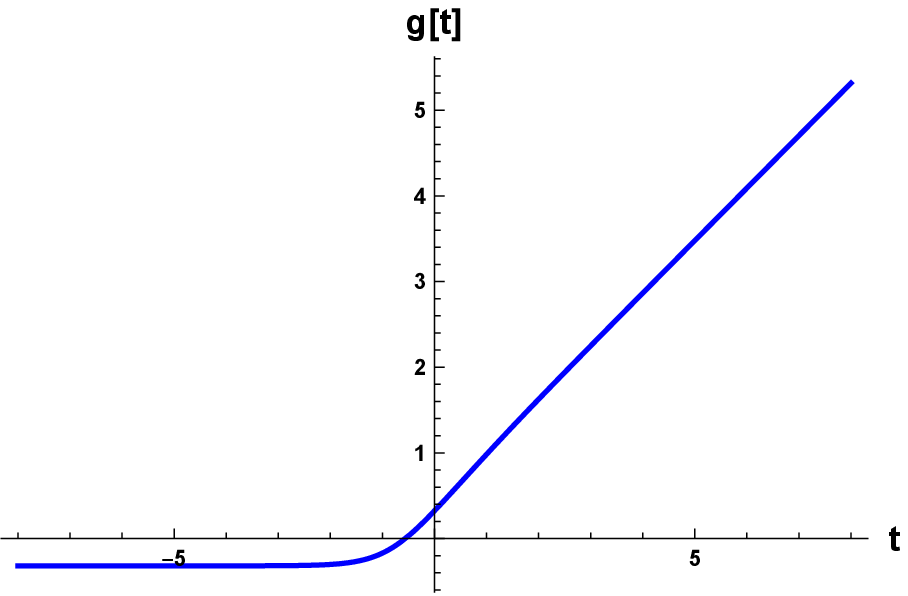}
\caption{Solution for $g$}  
  \end{subfigure}
      \begin{subfigure}[b]{0.3\textwidth}
    \includegraphics[width=\textwidth]{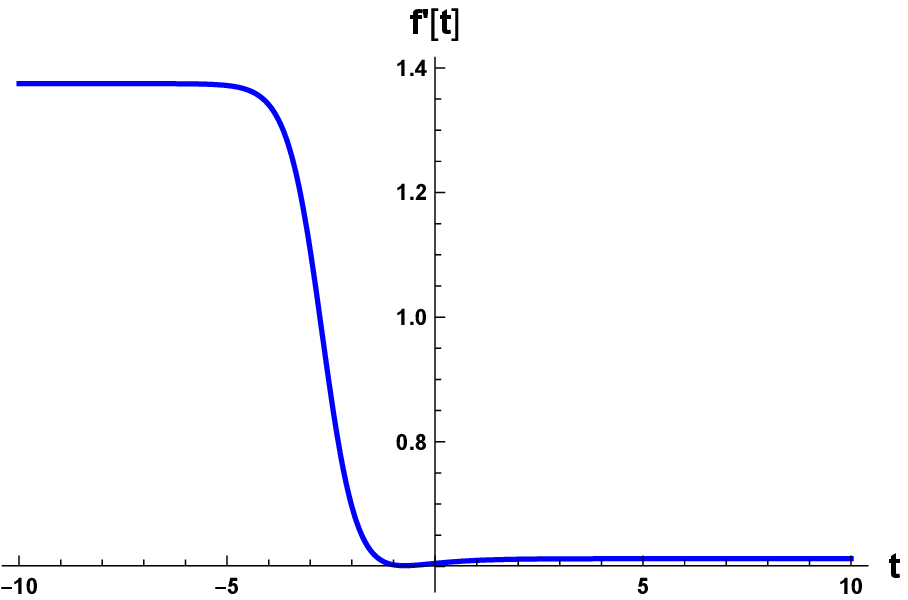}
\caption{Solution for $\dot f$}
    \end{subfigure}
  \begin{subfigure}[b]{0.3\textwidth}
    \includegraphics[width=\textwidth]{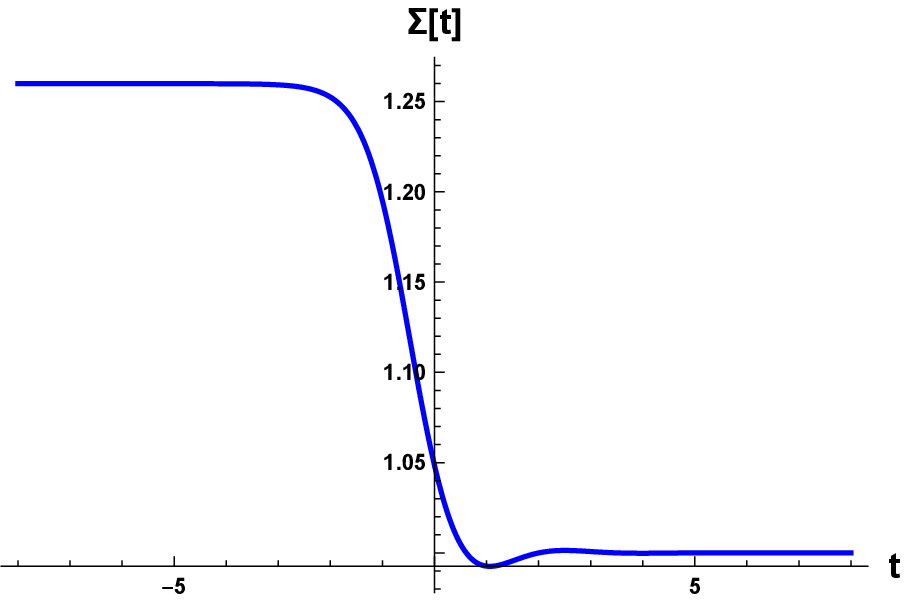}
\caption{Solution for $\Sigma$}
  \end{subfigure}
    \caption{A cosmological solution from $dS_3\times H^2$ (\ref{so31-dS3}) at the infinite past to the $dS_5$ solution (\ref{so31-dS5}) in the infinite future from $SO(1,1)\times SO(3,1)$ gauge group and $g_2 = 1$.}
    \label{fig:SO31-dS3-cf}
    \end{figure}

%%%%%%%%%%%%%%%%%%%%%%%%%%%%%%%%%%%%%%%%%%%%%%%%%%%%%%%%%%%
Numerically solving the equations (\ref{dS3eom}) with the potential (\ref{Vso41}) yields the following interpolating solution plotted in Fig. \ref{fig:SO31-dS3-cf}. 
\newpage
    %%%%%%%%%%%%%%%%%%%%%%%%%%%%%%%%%%%%%%%%%%%%%%%%%%%%%%%%%%%
\subsubsection{$dS_2\times \Sigma_3$}
%%%%%%%%%%%%%%%%%%%%%%%%%%%%%%%%%%%%%%%%%%%%%%%%%%%%%%%%%%%
This is the first gauge group under consideration that can give rise to $dS_2\times \S_3$ solutions because of the subgroup $SO(3)\subset SO(3,1)$. For this type of solutions, we turn on the full non-Abelian $SO(3)$ gauge field $A^{M=8}, A^{N=9}, A^{P=10}$ in (\ref{dS2-A-ans}) corresponding to generators $X_8, X_9, X_{10}$ of the $SO(3)$ factor. 
The gauge ansatz for $SO(3)$ is given by (\ref{dS2-A-ans}) and the resulting gauge field strengths by (\ref{HdS2}).
The condition (\ref{dS2twist}) in this case is
\beq
a = \frac{i}{g_2}\qquad (c =1). 
\label{dS2twist31}
\eeq
 The equations (\ref{dS2eom}) with the potential (\ref{Vso31}) yield the following $dS_2\times H^3$ solution. 
 \beq
\begin{array}{l}
\S_0= 2^{1/2} \zeta,\\\\
f_0= 2^{1/2}\, \sqrt{3} \, \zeta^2\,\,g_2,\\
\\
g_0= \dfrac{1}{2} \log \left\{4\left(\sqrt{1-\frac{3}{4} a^2 g_2^2}-1\right) \dfrac{\zeta^2}{3 g_2^2}\right\}
\\\\
\zeta = \lf[\dfrac{\frac{1}{2}a^2 g_2^2-\sqrt{1-\frac{3}{4} a^2 g_2^2}-1}{a^2 g_2^2} \rr]^{1/6}
\end{array} \label{so31-dS2-0}
\eeq
which becomes
\beq
\begin{array}{l}
\S_0= \sqrt[3]{2} \sqrt[6]{\sqrt{7}+3},\\\\
f_0= \sqrt[6]{2} \sqrt{3} \sqrt[3]{\sqrt{7}+3}\,\, g_2,
\\\\
g_0= \dfrac{1}{2} \log \left(\dfrac{2^{2/3} \left(\sqrt{7}-2\right) \sqrt[3]{\sqrt{7}+3}}{3 g_2^2}\right) \end{array} \label{so31-dS2}
\eeq
upon using (\ref{dS2twist31}).
%%%%%%%%%%%%%%%%%%%%%%%%%%%%%%%%%%%%%%%%%%%%%%%%%%%%%%%%%%%
\subsubsection{Cosmological solution between $dS_2\times \Sigma_3$ and $dS_5$}
\begin{figure}[!htb]
\centering 
  \begin{subfigure}[b]{0.3\textwidth}
    \includegraphics[width=\textwidth]{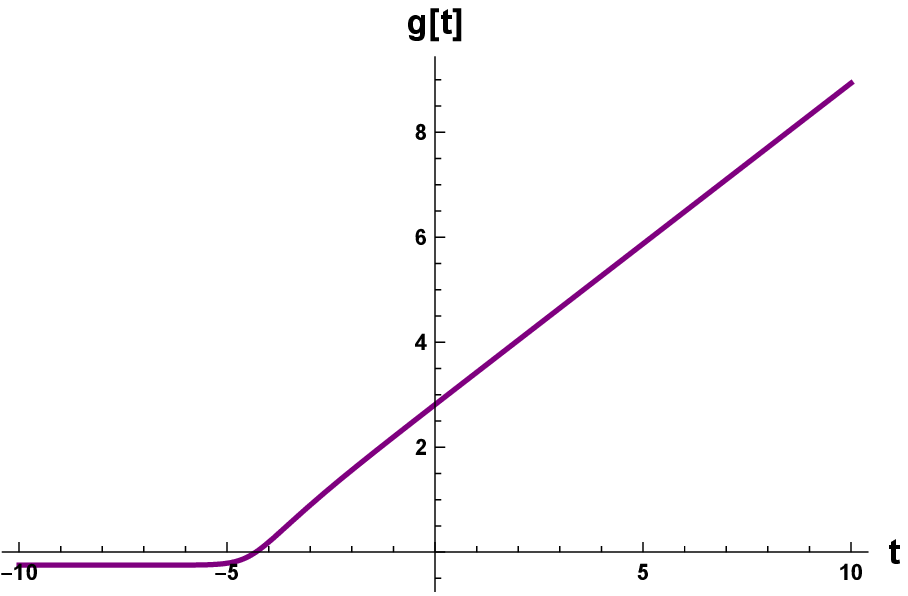}
\caption{Solution for $g$}  
  \end{subfigure}
      \begin{subfigure}[b]{0.3\textwidth}
    \includegraphics[width=\textwidth]{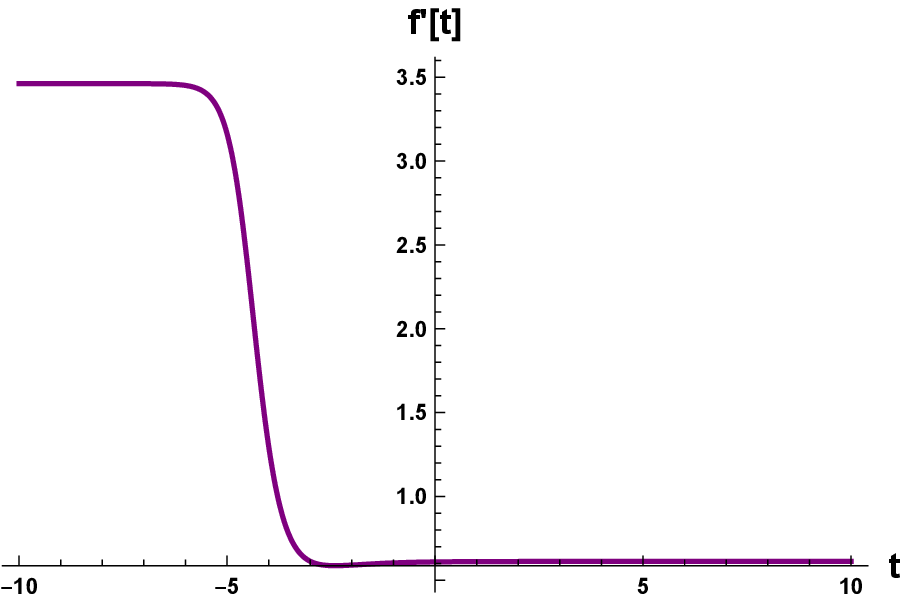}
\caption{Solution for $\dot f$}
    \end{subfigure}
      \begin{subfigure}[b]{0.3\textwidth}
    \includegraphics[width=\textwidth]{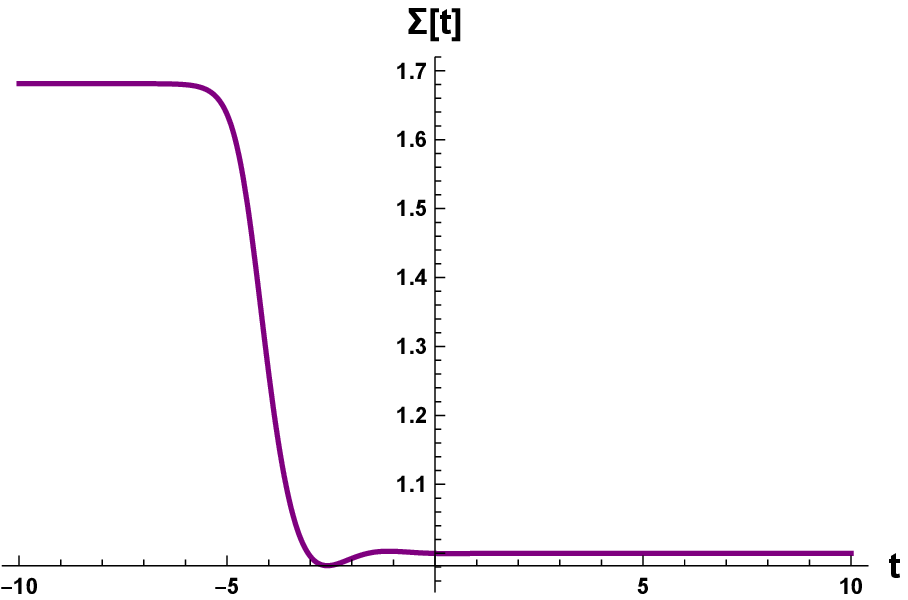}
\caption{Solution for $\Sigma$}
  \end{subfigure}
    \caption{A cosmological solution from $dS_3\times H^2$ (\ref{so31-dS2}) at the infinite past to the $dS_5$ (\ref{so31-dS5}) solution in the infinite future from $SO(1,1)\times SO(3,1)$ gauge group and $g_2 = 1$.}
    \label{fig:SO31-dS2-cf}
    \end{figure}

%%%%%%%%%%%%%%%%%%%%%%%%%%%%%%%%%%%%%%%%%%%%%%%%%%%%%%%%%%%
Numerically solving the equations (\ref{dS2eom}) with the potential (\ref{Vso41}) yields the following interpolating solution plotted in Fig. \ref{fig:SO31-dS2-cf}. 
    %\newpage
%%%%%%%%%%%%%%%%%%%%%%%%%%%%%%%%%%%%%%%%%%%%%%%%%%%%%%%%%%%
\subsection{$SO(1,1)^{(2)}_\text{diag}\times SO(3,1)$}
%%%%%%%%%%%%%%%%%%%%%%%%%%%%%%%%%%%%%%%%%%%%%%%%%%%%%%%%%%%
This gauged theory shares the same embedding tensor components for the $SO(3,1)$ factor as the $SO(1,1)\times SO(3,1)$ theory (see Table \ref{table:dS5}). Consequently, this particular theory has the same $dS_5$ solution (with different $g_1/g_2$ ratio) 
\beq
\S_0 = 1, \qquad f_0 = \frac{1}{2}\sqrt{\frac{3}{2}}g_2, \qquad g_1 =  \sqrt{\frac{3}{2}}\,g_2.\label{so312-dS5}
\eeq
 as the $SO(1,1)\times SO(3,1)$ theory (\ref{so31-dS5}), given by the scalar potential
\beq
V = \frac{3g_2^2 + g_1^2 \S^6}{2\S^2}. \label{Vso312}
\eeq
Furthermore, the $dS_3\times H^2$ and $dS_2\times H^3$ fixed point solutions of this theory are also identical to those of the  $SO(1,1)\times SO(3,1)$ theory which are given by (\ref{so31-dS3}, \ref{so31-dS3-0}) and (\ref{so31-dS2}, \ref{so31-dS2-0}), respectively. The cosmological solutions interpolating between these fixed point and the $dS_5$ solution (\ref{so312-dS5}) are  given in Fig. (\ref{fig:SO31-dS3-cf}) and Fig. (\ref{fig:SO31-dS2-cf}), respectively.
%%%%%%%%%%%%%%%%%%%%%%%%%%%%%%%%%%%%%%%%%%%%%%%%%%%%%%%%%%%
\subsection{$SO(1,1)\times SO(4,1)$}
%%%%%%%%%%%%%%%%%%%%%%%%%%%%%%%%%%%%%%%%%%%%%%%%%%%%%%%%%%%
The $SO(4,1)$-gauged theory is determined by the following scalar potential
\beq
V(\Sigma)&=& \frac{12g_2^2 + g_1^2\S^6}{4\S^2}, \label{Vso41}
\eeq
with the following $dS_5$ vacuum
\beq
\Sigma_0 = 1, \qquad f_0 = \frac{\sqrt{3}}{2}g_2,\qquad  g_2 = \sqrt{6} g_1.\label{so41-dS5}
\eeq
This gauged theory is different from the rest because it requires at least seven coupled vector multiplets (instead of five). This is in fact the largest gauging with $dS_5$ solution that one can have for $N=4$ 5D matter-coupled supergravity. The reason for this being that there are a maximum of only four non-compact directions that can be embedded in the R-symmetry group $SO(5)$, apart from the one taken by the $SO(1,1)$ factor. These four non-compact directions are all taken by the non-compact part of $SO(4,1)$. 
\\\indent
The embedding tensor for this theory is given in Table \ref{table:dS5}. The $SO(4)$ subgroup of the $SO(4,1)$ gauge factor is generated by $X_7, \ldots, X_{12}$. Equivalently, we can write
\beq
SO(4) \cong SO(3)_+ \times SO(3)_-, \label{41-33}
\eeq
which are generated by the following generators
\beq
SO(3)_\pm: && X_7 \pm X_{12}, \qquad X_8 \mp X_{11}, \qquad X_9 \pm X_{10}. \label{so3pm}
\eeq
%%%%%%%%%%%%%%%%%%%%%%%%%%%%%%%%%%%%%%%%%%%%%%%%%%%%%%%%%%%
\subsubsection{$dS_3\times \Sigma_2$}
%%%%%%%%%%%%%%%%%%%%%%%%%%%%%%%%%%%%%%%%%%%%%%%%%%%%%%%%%%%
For this solution, we turn on the gauge field corresponding to $SO(2)\subset SO(3)_+\subset SO(4,1)$ generated by $X_7+ X_{12}$ in (\ref{so3pm}).  An equivalent choice is the gauge field corresponding to $SO(2)\subset SO(3)_-$ generated by $X_7-X_{12}$. 
The gauge field ansatz is slightly different from (\ref{dS3-A-ans}). For $\S_2 = S^2$,
\beq
A^7_\phi = A^{12}_\phi = \frac{a}{\sqrt{2}}\cos\theta. \label{so41-A}
\eeq
with the resulting gauge field strengths
\beq
\mc H^7_{\theta\phi} = \mc H^{12}_{\theta\phi} = \frac{a}{\sqrt{2}}\sin\theta.\label{so41-H}
\eeq
For $\S_2 = H^2$, one simply replaces $\cos\theta$ in (\ref{so41-A}) with $\cosh\theta$ and $\sin\theta$ in (\ref{so41-H}) with $\sinh\theta$. The resulting equations of motion with the above gauge ansatz are given by (\ref{AdS3eom}).
With the specific scalar potential (\ref{Vso41}), the $dS_3$ equations (\ref{dS3eom}) yield the following $dS_3\times H^2$ fixed-point solution
\beq
\begin{array}{l}
\Sigma_0= 3^{-1/6}\,\zeta,
\\
f_0=\dfrac{3^{1/6}}{\sqrt{2}}\, g_2 \,\zeta^2,
\\
g_0= \dfrac{1}{2} \log \left(\dfrac{3^{-1/3}}{6 }\left(\sqrt{1-6 a^2 g_2^2}-1\right) \,\dfrac{\zeta^2}{g_2^2}\right) 
\\
\zeta = \lf[\dfrac{6 a^2 g_2^2-\sqrt{1-6 a^2 g_2^2}-1}{a^2 g_2^2}\rr]^{1/6}\end{array}
.  \label{so41-dS3-0}
\eeq
The solution is real if we impose (\ref{dS3twist})
\beq
\Sigma_0&=& \sqrt[6]{\frac{1}{3} \left(\sqrt{7}+7\right)}, \non
f_0&=& \frac{\sqrt[6]{3} \sqrt[3]{\sqrt{7}+7} }{\sqrt{2}}\,g_2,\non
g_0&=& \frac{1}{2} \log \left(\frac{\left(\sqrt{7}-1\right) \sqrt[3]{\frac{1}{3} \left(\sqrt{7}+7\right)}}{6 g_2^2}\right)\,\,. \label{so41-dS3}
\eeq
%%%%%%%%%%%%%%%%%%%%%%%%%%%%%%%%%%%%%%%%%%%%%%%%%%%%%%%%%%%
\subsubsection{Cosmological solution between $dS_3\times \Sigma_2$ and $dS_5$}
\begin{figure}[!htb]
\centering 
  \begin{subfigure}[b]{0.3\textwidth}
    \includegraphics[width=\textwidth]{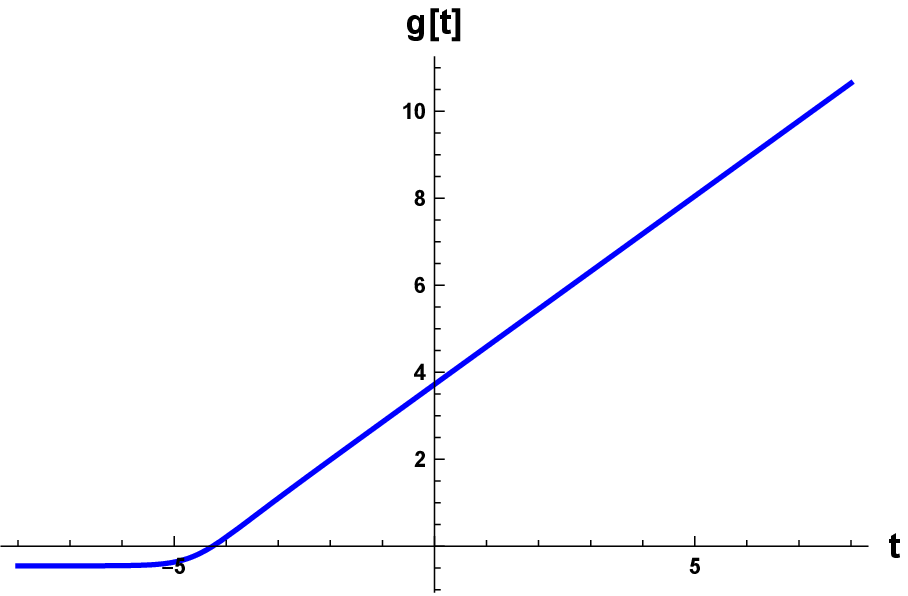}
\caption{Solution for $g$}  
  \end{subfigure}
      \begin{subfigure}[b]{0.3\textwidth}
    \includegraphics[width=\textwidth]{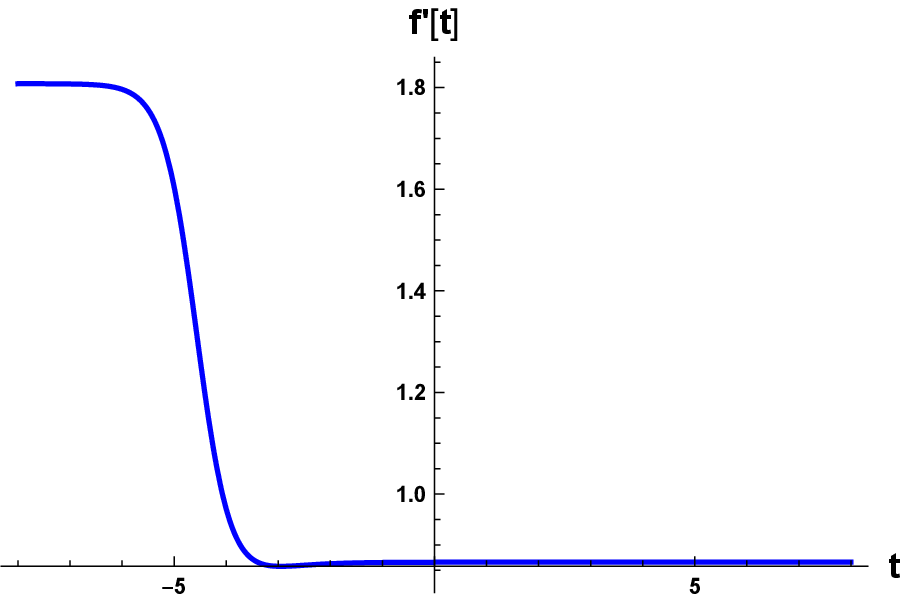}
\caption{Solution for $\dot f$}
    \end{subfigure}
    \begin{subfigure}[b]{0.3\textwidth}
    \includegraphics[width=\textwidth]{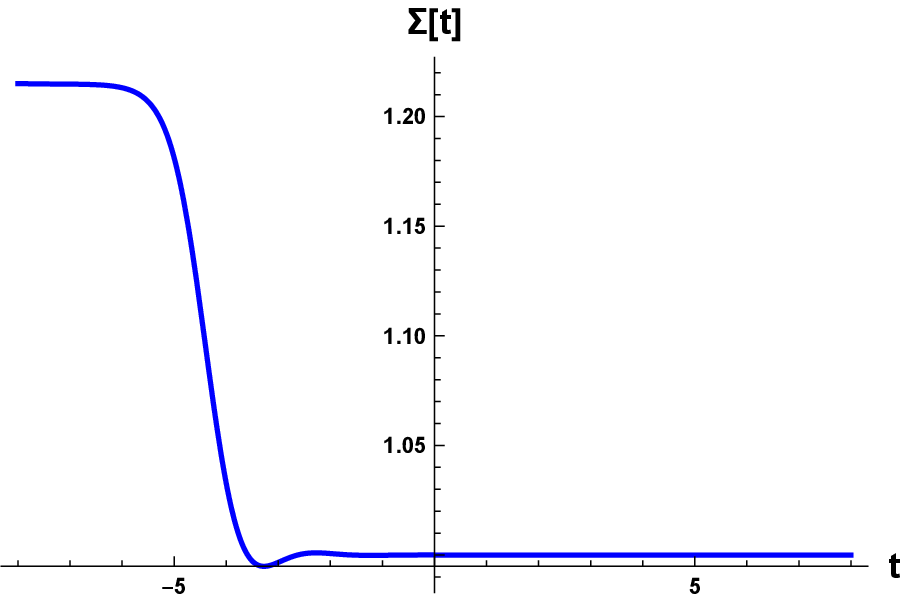}
\caption{Solution for $\Sigma$}
  \end{subfigure}
    \caption{A cosmological solution from $dS_3\times H^2$ (\ref{so41-dS3}) at the infinite past to the $dS_5$ solution (\ref{so41-dS5}) in the infinite future from $SO(1,1)\times SO(4,1)$ gauge group and $g_2 = 1$.}
    \label{fig:SO41-dS3-cf}
    \end{figure}

%%%%%%%%%%%%%%%%%%%%%%%%%%%%%%%%%%%%%%%%%%%%%%%%%%%%%%%%%%%
We now numerically solve (\ref{dS3eom}) with the potential (\ref{Vso41}) to obtain the following interpolating solution plotted in Fig.\ref{fig:SO41-dS3-cf}.
%\newpage
%%%%%%%%%%%%%%%%%%%%%%%%%%%%%%%%%%%%%%%%%%%%%%%%%%%%%%%%%%%
\subsubsection{$dS_2\times \Sigma_3$ }
%%%%%%%%%%%%%%%%%%%%%%%%%%%%%%%%%%%%%%%%%%%%%%%%%%%%%%%%%%%
In this case, we can either turn on the $SO(3)_+$ or $SO(3)_-$ subgroup of $SO(4)\in SO(4,1)$ (\ref{so3pm}). With both choices being equivalent, we will turn on the gauge fields corresponding to the generators 
$(X_7+X_{12}), (X_{8} - X_{11}), (X_9+X_{10})$ of
$SO(3)_+$. Since the ansatz is slightly different than (\ref{dS2-A-ans}), we will explicitly specify it here for the case of $\S_3 = S^3$,
\beq
A^7_\phi = A^{12}_5 &=& a_7 \,\cos\psi \sin\theta \non
A^8_\phi = -A^{11}_5 &=& a_8 \cos\theta \non
A^9_\theta = A^{10}_\theta &=& a_9 \cos\psi.  \label{so41-A-ans}
\eeq
The resulting field strengths are
\beq
\mc H^7_{\psi\phi} = \mc H^{12}_{\psi\theta} &=& \frac{a}{\sqrt{2}}\sin\theta \sin\psi \non
\mc H^8_{\theta\phi} = -\mc H^{11}_{\theta\phi} &=& \frac{a}{\sqrt{2}}\sin\theta \sin^2\psi \non
\mc H^9_{\psi\theta} = \mc H^{10}_{\psi\theta} &=& \frac{a}{\sqrt{2}}\sin\psi \label{so41-H}
\eeq
after setting $a_7 = a_8 = a_9 = \frac{a}{\sqrt{2}}$ and using (\ref{dS2twist})
\beq
g_2 = \frac{i}{ \sqrt{2}}\,a, \qquad \lf( c =\sqrt{2}\rr). \label{dS2twist41}
\eeq
For $\S_3 = H^3$, the gauge ansatz and resulting field strengths are given by (\ref{so41-A-ans}) and (\ref{so41-H}), respectively, but with $\cos \psi$ replaced by $\cosh\psi$ and $\sin \psi$ replaced by $\sinh\psi$. 
The equations (\ref{dS2eom}) with the scalar potential (\ref{Vso41}) yield the following $dS_2\times H^3$ fixed point solution
\beq
\begin{array}{l}
\S_0= \sqrt[3]{2} \,\zeta
\\\\
f_0= \sqrt{3}\,\,\zeta^2\, g_2 ,\\\\
g_0= \dfrac{1}{2} \log \left\{2\left(\sqrt{1-\frac{3}{2} a^2 g_2^2}-1\right) \dfrac{\zeta^2}{6 g_2^2}\right\}
\\\\
\zeta =\lf[\dfrac{a^2 g_2^2-\left(\sqrt{1-\frac{3}{2} a^2 g_2^2}+1\right)}{a^2 g_2^2}\rr]^{1/6} 
\end{array}\label{so41-dS2-0}
\eeq
which becomes
\beq
\begin{array}{l}
\S_0= \sqrt[3]{2} \sqrt[6]{ \sqrt{7}+3},\\\\
f_0= \sqrt{3} \sqrt[3]{4 \left(\sqrt{7}+2\right)+4}\,\, g_2,
\\\\
g_0= \dfrac{1}{2} \log \left(\dfrac{\left(\sqrt{7}-2\right) \sqrt[3]{\frac{1}{2} \left(\sqrt{7}+3\right)}}{3 g_2^2}\right) 
\end{array}\label{so41-dS2}
\eeq
after using (\ref{dS2twist41}).
%%%%%%%%%%%%%%%%%%%%%%%%%%%%%%%%%%%%%%%%%%%%%%%%%%%%%%%%%%%
\subsubsection{Cosmological solution between $dS_2\times \Sigma_3$ and $dS_5$}
\begin{figure}[!htb]
\centering 
  \begin{subfigure}[b]{0.3\textwidth}
    \includegraphics[width=\textwidth]{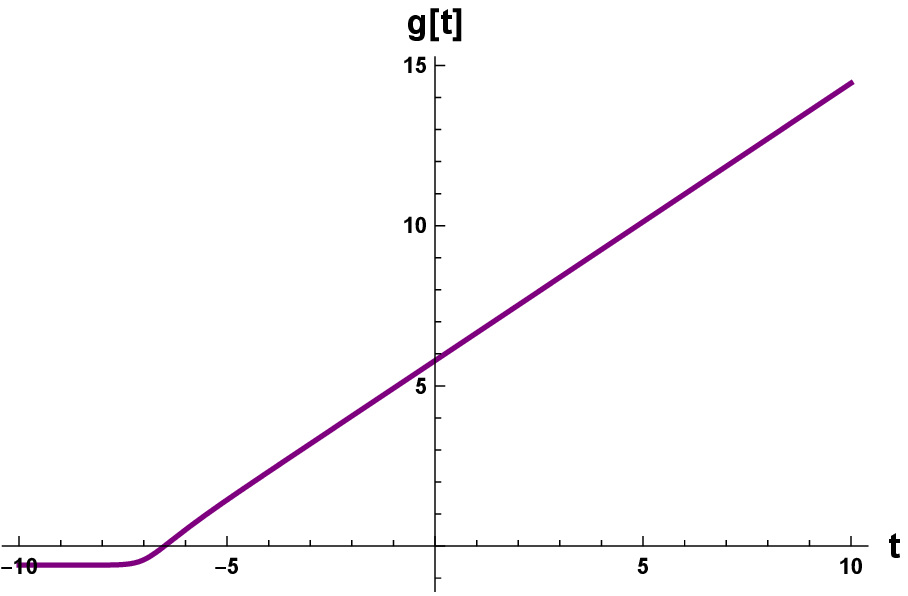}
\caption{Solution for $g$}  
  \end{subfigure}
      \begin{subfigure}[b]{0.3\textwidth}
    \includegraphics[width=\textwidth]{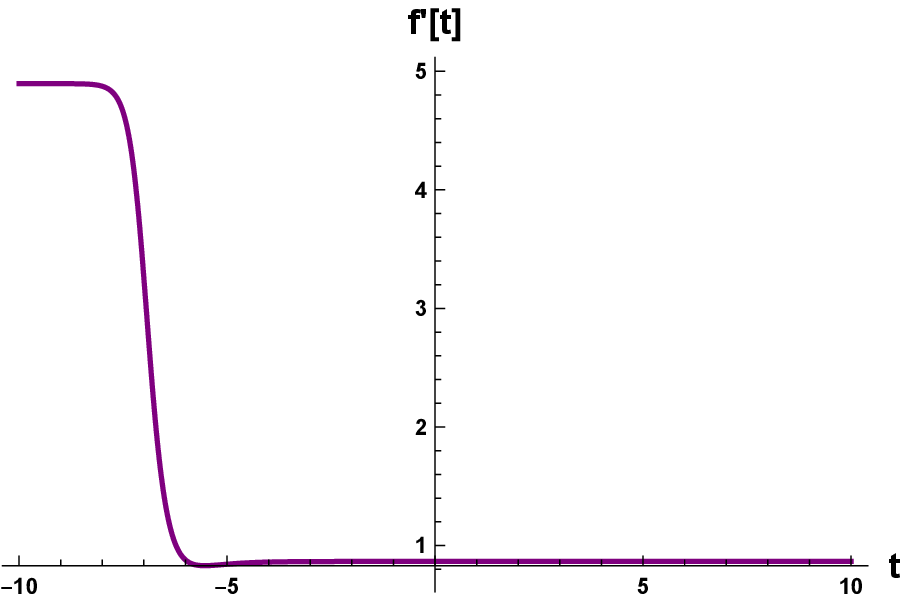}
\caption{Solution for $\dot f$}
    \end{subfigure}
      \begin{subfigure}[b]{0.3\textwidth}
    \includegraphics[width=\textwidth]{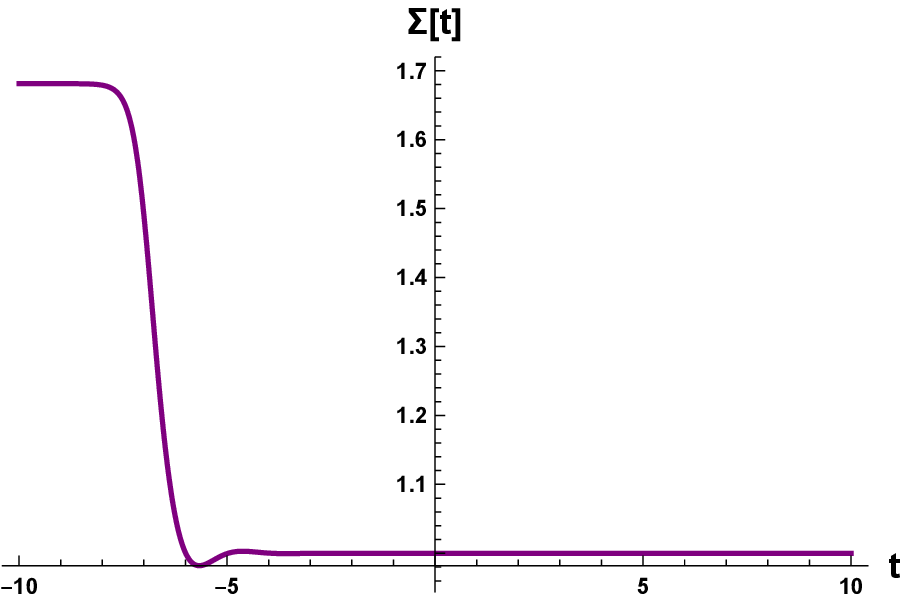}
\caption{Solution for $\Sigma$}
  \end{subfigure}
    \caption{A cosmological solution from $dS_2\times H^3$  (\ref{so41-dS2}) at the infinite past to the $dS_5$ (\ref{so41-dS5}) solution in the infinite future from $SO(1,1)\times SO(4,1)$ gauge group and $g_2 = 1$.}
    \label{fig:SO41-dS2-cf}
    \end{figure}

%%%%%%%%%%%%%%%%%%%%%%%%%%%%%%%%%%%%%%%%%%%%%%%%%%%%%%%%%%%
We now numerically solve (\ref{dS2eom}) with the potential (\ref{Vso41}) and obtain the following interpolating solution plotted in Fig.\ref{fig:SO41-dS2-cf}.
    \newpage 
   %\clearpage
%%%%%%%%%%%%%%%%%%%%%%%%%%%%%%%%%%%%%%%%%%%%%%%%%%%%%%%%%%%
\subsection{$SO(1,1)\times SU(2,1)$}
%%%%%%%%%%%%%%%%%%%%%%%%%%%%%%%%%%%%%%%%%%%%%%%%%%%%%%%%%%%
The scalar potential of this theory is
\beq
V= \frac{24 g_2^2 + g_1^2 \S^6}{4\S^2}\label{Vsu21}
\eeq
with the following $dS_5$ vacuum
\beq
\S_0 = 1,\qquad f_0 =\sqrt{\frac{3}{2}}g_2, \qquad g_1 =2\sqrt{3}g_2.\label{su21-dS5}
\eeq
The embedding tensor for this theory is given in Table \ref{table:dS5}.
%%%%%%%%%%%%%%%%%%%%%%%%%%%%%%%%%%%%%%%%%%%%%%%%%%%%%%%%%%%
\subsubsection{$dS_3\times \Sigma_2$ solution }
%%%%%%%%%%%%%%%%%%%%%%%%%%%%%%%%%%%%%%%%%%%%%%%%%%%%%%%%%%%
To obtain this solution, we turn on the $U(1)$ gauge field $A^9$ or $A^{10}$ (\ref{dS3-A-ans}) which corresponds to either $U(1)\subset SU(2)\subset SU(2,1)$ or $U(1)\subset SU(2,1)$, respectively. Both choices lead to the same result (\ref{dS3eom}). 
The equations (\ref{dS3eom}) together with the scalar potential (\ref{Vsu21}) admit the following $dS_3\times H^2$ fixed point solution
\beq
\begin{array}{l}
\S_0=6^{-1/6}\,\zeta,\\
f_0 = \lf(\dfrac{\sqrt{3}}{2}\rr)^{1/3}\,g_2\,\zeta^2,
\\
g_0= \dfrac{1}{2} \log \left(\dfrac{6^{-1/3}}{12}\left(\sqrt{1-12 a^2 g_2^2}-1\right)\,\dfrac{\zeta^2}{ g_2^2}\right) 
\\
\zeta =\lf[\dfrac{12 a^2 g_2^2-\sqrt{1-12 a^2 g_2^2}-1}{a^2 g_2^2}\rr]^{1/6}\end{array} \label{su21-dS3-0}
\eeq
which is real and becomes
\beq
\S_0&=& \sqrt[6]{\frac{1}{6} \left(\sqrt{13}+13\right)},\non
f_0&=& \sqrt[6]{3} \sqrt[3]{\frac{1}{2} \left(\sqrt{13}+13\right)} g_2,
\non
g_0&=& \frac{1}{2} \log \left(\frac{\left(\sqrt{13}-1\right) \sqrt[3]{\frac{1}{6} \left(\sqrt{13}+13\right)}}{12 g_2^2}\right)
\label{su21-dS3}
\eeq
after using the condition (\ref{dS3twist}).
%%%%%%%%%%%%%%%%%%%%%%%%%%%%%%%%%%%%%%%%%%%%%%%%%%%%%%%%%%%
\subsubsection{Cosmological solution between $dS_3\times \Sigma_2$ and $dS_5$}
%%%%%%%%%%%%%%%%%%%%%%%%%%%%%%%%%%%%%%%%%%%%%%%%%%%%%%%%%%%
The interpolating solution between (\ref{su21-dS3}) and (\ref{su21-dS5}), plotted in Fig.\ref{fig:SU21-dS3-cf} below, is obtained by numerically solving (\ref{dS2eom}) with (\ref{Vsu21}).
\begin{figure}[!htb]
\centering 
  \begin{subfigure}[b]{0.3\textwidth}
    \includegraphics[width=\textwidth]{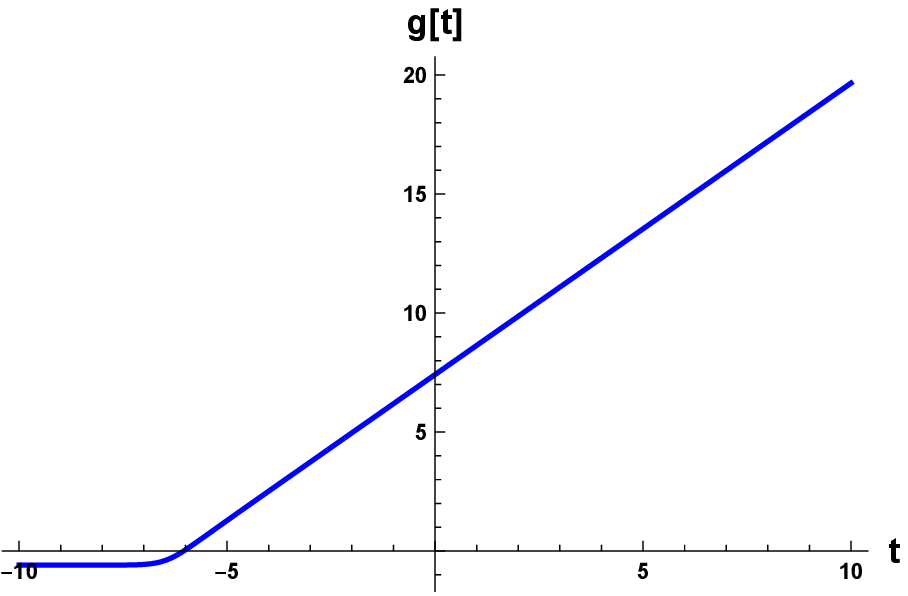}
\caption{Solution for $g$}  
  \end{subfigure}
      \begin{subfigure}[b]{0.3\textwidth}
    \includegraphics[width=\textwidth]{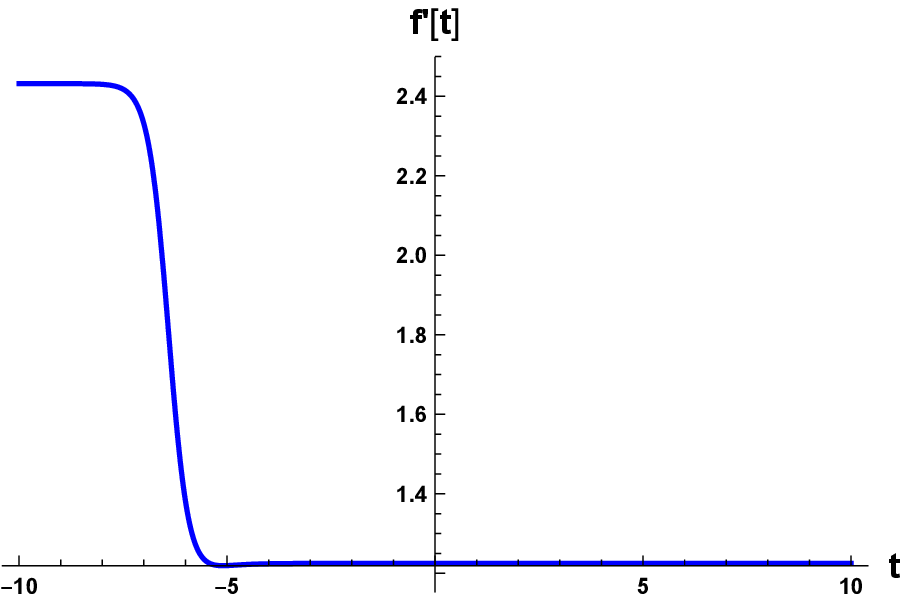}
\caption{Solution for $\dot f$}
    \end{subfigure}
      \begin{subfigure}[b]{0.3\textwidth}
    \includegraphics[width=\textwidth]{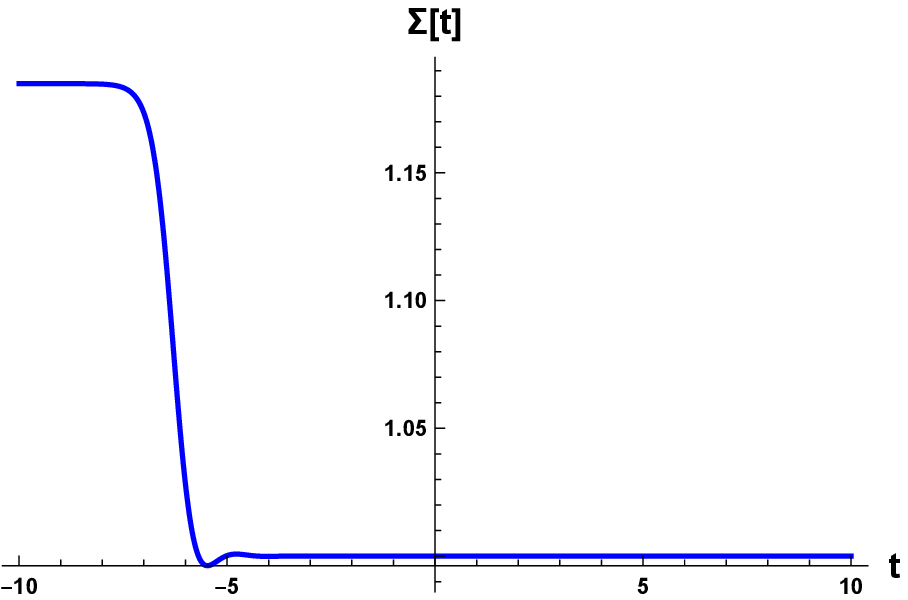}
\caption{Solution for $\Sigma$}
  \end{subfigure}
    \caption{A cosmological solution from $dS_3\times H^2$ (\ref{su21-dS3}) at the infinite past to the $dS_5$ solution (\ref{su21-dS5}) in the infinite future from $SO(1,1)\times SU(2,1)$ gauge group and $g_2 = 1$.}
    \label{fig:SU21-dS3-cf}
    \end{figure}
   %\newpage
  
%%%%%%%%%%%%%%%%%%%%%%%%%%%%%%%%%%%%%%%%%%%%%%%%%%%%%%%%%%%
\subsubsection{$dS_2\times \Sigma_3$}
%%%%%%%%%%%%%%%%%%%%%%%%%%%%%%%%%%%%%%%%%%%%%%%%%%%%%%%%%%%
For this solution, we need to turn on the gauge fields $A^{M=7}, A^{N=8}, A^{P=9}$ (\ref{dS2-A-ans}) corresponding to the gauge generators $X_7, X_8, X_9$ of $SU(2)\subset SU(2,1)$. The gauge ansatz for $SO(3)$ is given by (\ref{dS2-A-ans} and the resulting gauge field strengths by (\ref{HdS2}).
The condition (\ref{dS2twist}) in this case is
\beq
 g_2 = \frac{i}{2a}, \qquad (c=2). \label{dS2twist21}
\eeq
The equations (\ref{dS2eom}) with the scalar potential (\ref{Vsu21}) yield the following $dS_2\times H^3$ fixed point solution
\beq
\begin{array}{l}
\S_0= \sqrt[6]{2} \,\zeta\\
\\
f_0= 2^{5/6} \sqrt{3} \,\zeta^2\,\,g_2,\\
\\
g_0= \dfrac{1}{2} \log \left\{\left(\sqrt{1-3 a^2 g_2^2}-1\right) \dfrac{\zeta^2}{3\ 2^{2/3} g_2^2}\right\}
\\\\
\zeta = \lf[\dfrac{2 a^2 g_2^2-\sqrt{1-3 a^2 g_2^2}-1}{a^2 g_2^2}\rr]^{1/6},
\end{array}\label{su21-dS2-0}
\eeq
which becomes
\beq
\begin{array}{l}
\S_0= \sqrt[3]{2} \sqrt[6]{\sqrt{7}+3},
\\\\
f_0= 2 \sqrt[6]{2} \sqrt{3} \sqrt[3]{\sqrt{7}+3} \,\,g_2,
\\\\
g_0= \dfrac{1}{2} \log \left(\dfrac{\left(\sqrt{7}-2\right) \sqrt[3]{\frac{1}{2} \left(\sqrt{7}+3\right)}}{6 g_2^2}\right)
\end{array} \label{su21-dS2}
\eeq
after imposing the condition (\ref{dS2twist21}). 
%%%%%%%%%%%%%%%%%%%%%%%%%%%%%%%%%%%%%%%%%%%%%%%%%%%%%%%%%%%
\subsubsection{Cosmological solution between $dS_2\times \Sigma_3$ and $dS_5$}
\begin{figure}[!htb]
\centering 
  \begin{subfigure}[b]{0.3\textwidth}
    \includegraphics[width=\textwidth]{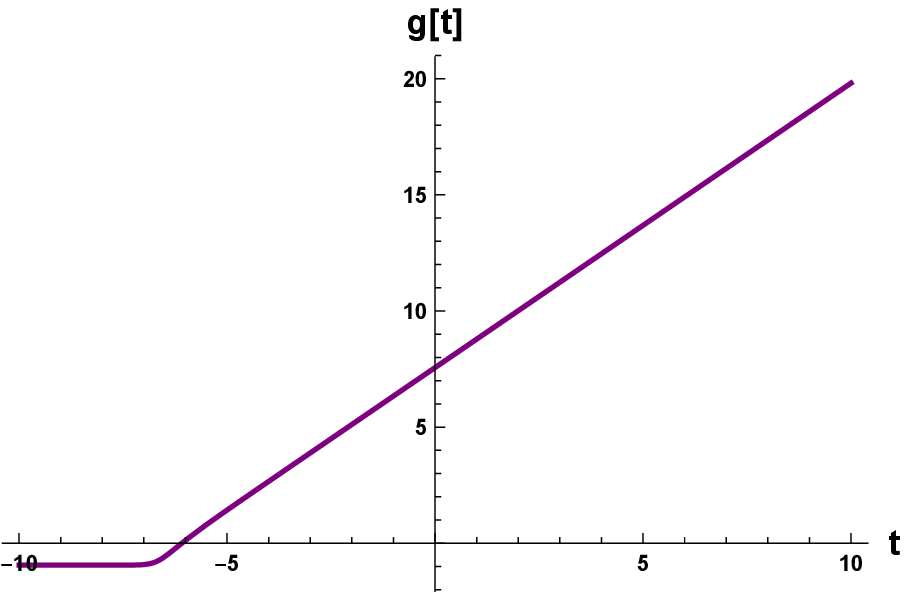}
\caption{Solution for $g$}  
  \end{subfigure}
      \begin{subfigure}[b]{0.3\textwidth}
    \includegraphics[width=\textwidth]{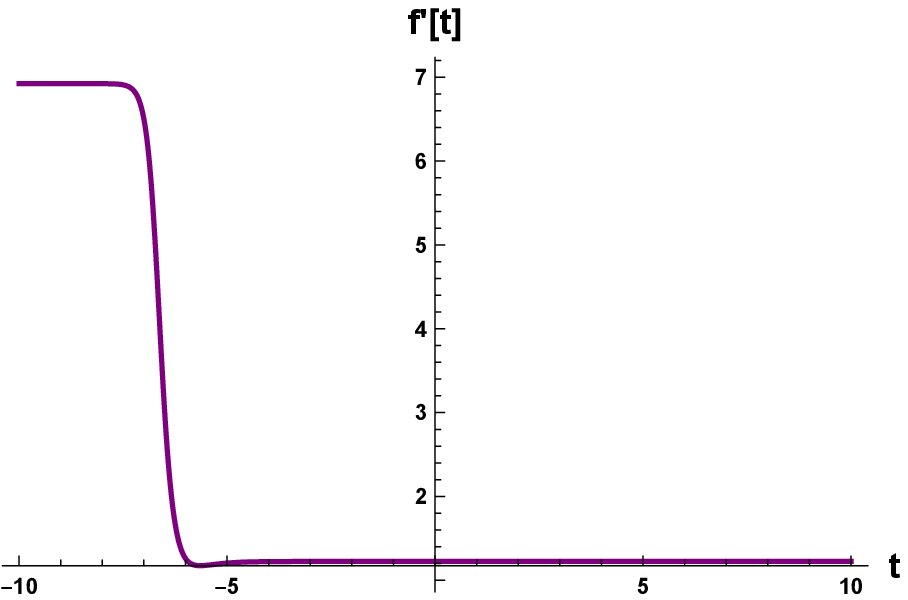}
\caption{Solution for $\dot f$}
    \end{subfigure}
      \begin{subfigure}[b]{0.3\textwidth}
    \includegraphics[width=\textwidth]{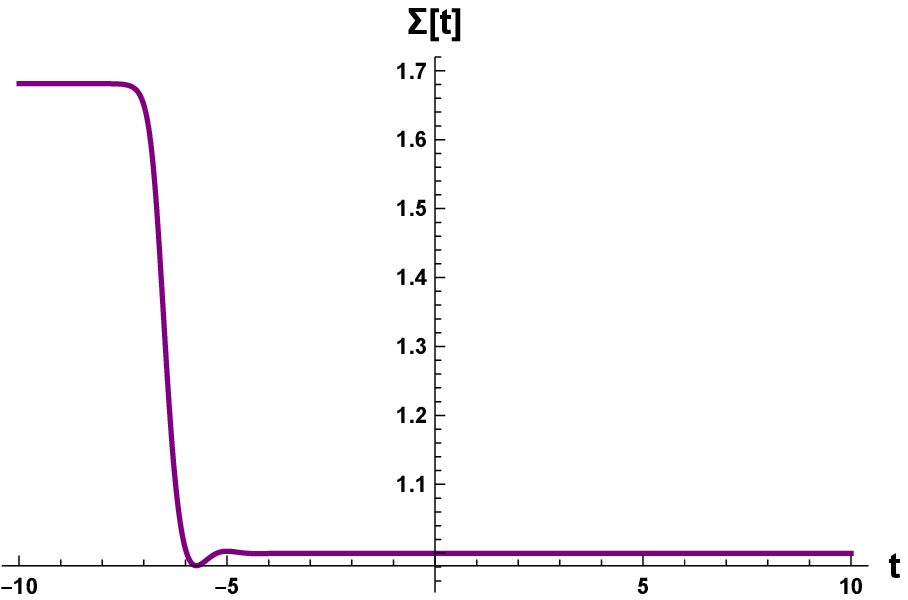}
\caption{Solution for $\Sigma$}
  \end{subfigure}
    \caption{A cosmological solution from $dS_3\times H^2$ (\ref{su21-dS2}) at the infinite past to the $dS_5$ (\ref{su21-dS5}) solution in the infinite future from $SO(1,1)\times SU(2,1)$ gauge group and $g_2 = 1$.}
    \label{fig:SU21-dS2-cf}
    \end{figure}

%%%%%%%%%%%%%%%%%%%%%%%%%%%%%%%%%%%%%%%%%%%%%%%%%%%%%%%%%%%
The interpolating solution between (\ref{su21-dS2}) and (\ref{su21-dS5}), plotted in Fig. \ref{fig:SU21-dS2-cf}, is obtained by numerically solving (\ref{dS2eom}) with (\ref{Vsu21}).
   %\newpage
   \clearpage
    %%%%%%%%%%%%%%%%%%%%%%%%%%%%%%%%%%%%%%%%%%%%%%%%%%%%%%%%%%%
\subsection{Summary of all solutions}
\begin{figure}[!htb]
\centering 
  \begin{subfigure}[b]{0.5\textwidth}
    \includegraphics[width=\textwidth]{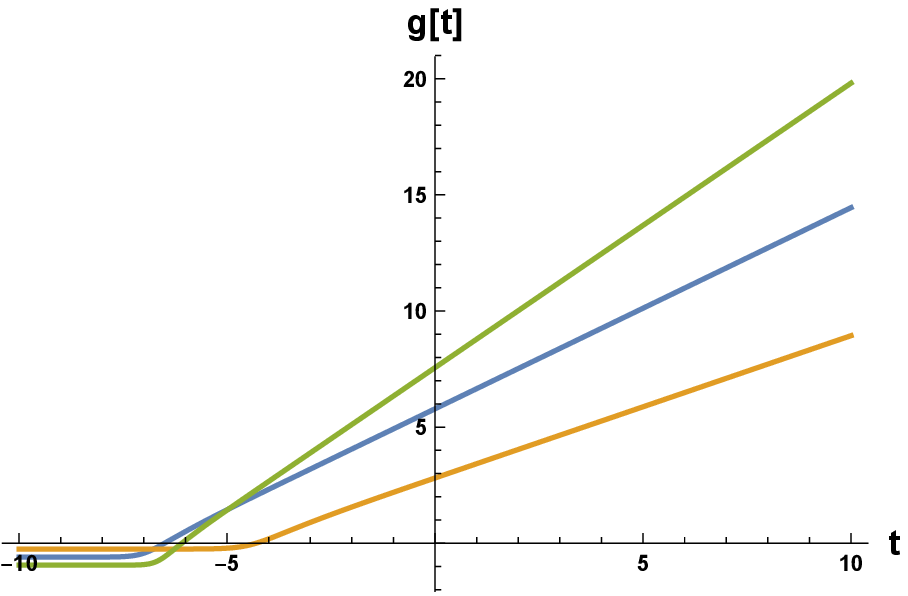}
\caption{Solution for $g$}  
  \end{subfigure}
      \begin{subfigure}[b]{0.5\textwidth}
    \includegraphics[width=\textwidth]{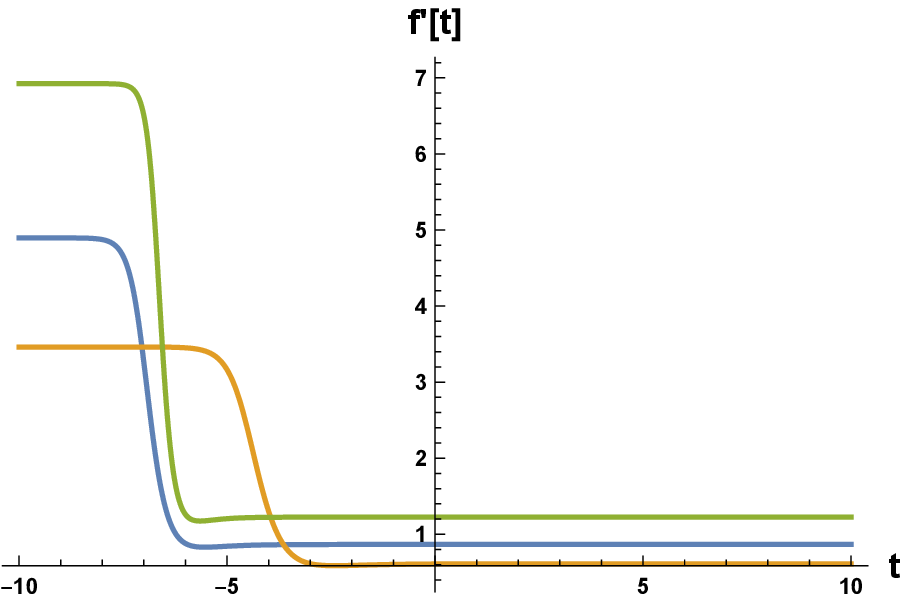}
\caption{Solution for $\dot f$}
  \begin{subfigure}[b]{1.35\textwidth}
    \includegraphics[width=\textwidth]{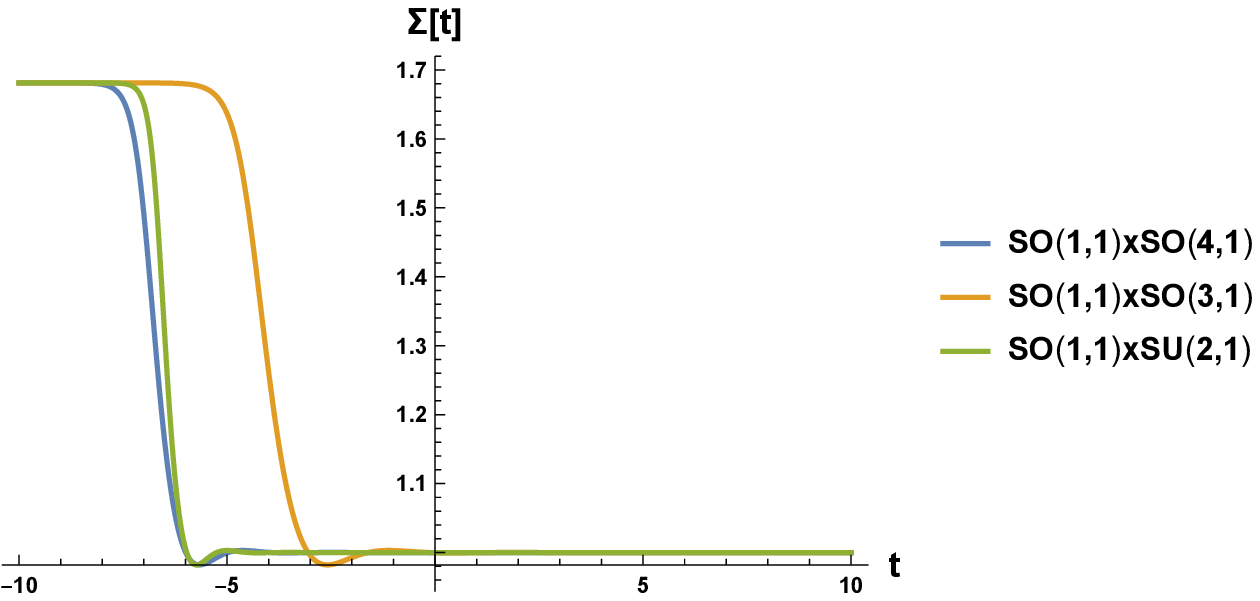}
\caption{Solution for $\Sigma$}
  \end{subfigure}  \end{subfigure}
    \caption{All cosmological solutions from $dS_2\times H^3$ at the infinite past to the $dS_5$ solution in the infinite future from all gauge groups and $g_2 = 1$. Note that the solution of the $SO(1,1)^{(2)}_\text{diag}\times SO(3,1)$ theory is the same as that of the $SO(1,1)\times SO(3,1)$.}
    \label{fig:all-dS2-cf}
    \end{figure}

\begin{figure}[!htb]
\centering 
  \begin{subfigure}[b]{0.5\textwidth}
    \includegraphics[width=\textwidth]{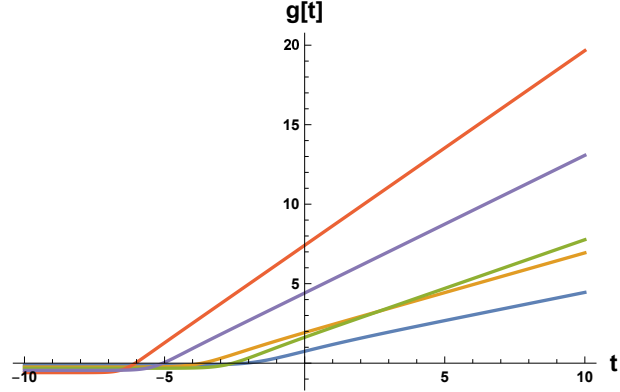}
\caption{Solution for $g$}  
  \end{subfigure}
      \begin{subfigure}[b]{0.5\textwidth}
    \includegraphics[width=\textwidth]{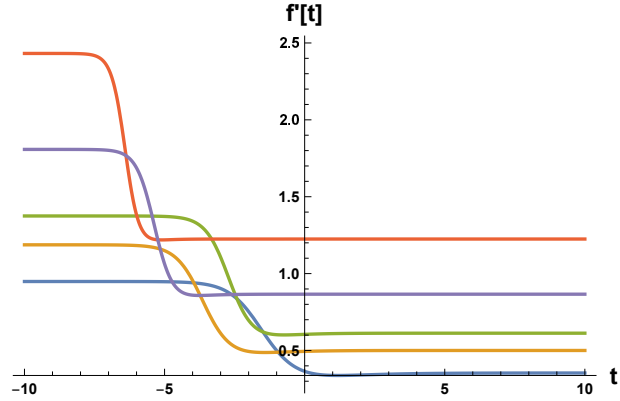}
\caption{Solution for $\dot f$}
  \begin{subfigure}[b]{1.35\textwidth}
    \includegraphics[width=\textwidth]{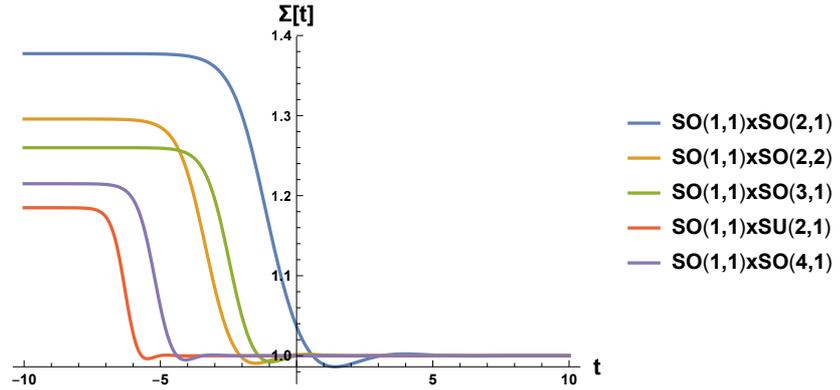}
\caption{Solution for $\Sigma$}
    \end{subfigure}  \end{subfigure}
    \caption{All cosmological solutions from $dS_3\times H^2$ at the infinite past to the $dS_5$ solution in the infinite future from all gauge groups and $g_2 = 1$. The solutions of the $SO(1,1)^{(n)}_\text{diag}\times SO(2,1)$ ($n=2,3$) and $SO(1,1)^{(2)}_\text{diag}\times SO(3,1)$ theories are given by those of $SO(1,1)\times SO(2,1)$ and $SO(1,1)\times SO(3,1)$ theories, respectively.}
    \label{fig:all-dS3-cf}
    \end{figure}

%%%%%%%%%%%%%%%%%%%%%%%%%%%%%%%%%%%%%%%%%%%%%%%%%%%%%%%%%%%
In this section, we summarize all the solutions listed above. 
Given the $g_1/g_2$ ratios in Table \ref{table:dS5-V}, the $dS_{3,2}\times H_{2,3}$ fixed point solutions of those gauge groups without the $SO(1,1)_\text{diag}$ factor can be rewritten by replacing $g_2$ with $g_1$\footnote{except in the denominators of $\zeta_{dS_3}$ in (\ref{all-dS3}) and $\zeta_{dS_2}$ in (\ref{all-dS2})} so that they assume the following some common form. 
\\\indent
For $dS_3\times H^2$ solution, equations (\ref{so21-dS3-0}, \ref{so22-dS3-0}, \ref{so31-dS3-0}, \ref{so41-dS3-0}, \ref{su21-dS3-0}) can be rewritten as
\beq
dS_3\times H^2: \qquad \begin{array}{l} 
\S_0 = c_1 \, \zeta_{dS_3} \\\\
 f_0 = c_2\, g_1\,\zeta_{dS_3}^2\\\\
 g_0 = \dfrac{1}{2}\log\lf\{c_3\, \lf(\sqrt{1-a^2 g_1^2}-1\rr) \dfrac{\zeta_{dS_3}^2}{g_1^2}\rr\}\\
\\
\zeta_{dS_3} = \lf[\dfrac{a^2 g_1^2 - \sqrt{1-a^2g_1^2} -1}{a^2 g_2^2}\rr]^{1/6},\end{array} \,\,.\label{all-dS3}
\eeq
For $dS_2\times H^3$ solution, equations ( \ref{so31-dS2-0}, \ref{so41-dS2-0}, \ref{su21-dS2-0}) can be rewritten as
\beq 
dS_2\times H^3:  \qquad 
\begin{array}{l}
\S_0 = c'_1 \zeta_{dS_2}\\\\
f_0 = c'_2 \,\zeta_{dS_2}^2\,g_1 \\\\
g_0 = \dfrac{1}{2}\log\lf\{ c'_3 \lf(\sqrt{1-\frac{1}{4}a^2 g_1^2} \rr) \dfrac{\zeta_{dS_2}^2}{g_1^2}\rr\}\\
\\
\zeta_{dS_2} = \lf[\dfrac{\frac{1}{6} a^2 g_1^2-\sqrt{1-\frac{1}{4} a^2 g_1^2}-1}{a^2 g_2^2}\rr]^{1/6},
\end{array} \label{all-dS2}
\eeq
\indent
For those gauge groups containing the $SO(1,1)^{(d)}_\text{diag}$ ($d=2,3)$ factor, the solutions in terms of $g_2$ are identical to those in the gauge groups $SO(1,1)\times SO(d,1)$ ($d=2,3)$, but when rewritten in terms of $g_1$\footnote{except in the denominators of $\zeta^{(2)}_{dS_3}$ in \ref{dS3-so112}, $\zeta^{(3)}_{dS_3}$ in \ref{dS3-so213}, and $\zeta^{(2)}_{dS_2}$ in \ref{dS2-so312}}, the solutions assume different forms than those given in (\ref{all-dS3}, \ref{all-dS2}). Specifically, the solutions (\ref{so21-dS3-0}) and (\ref{so31-dS3-0}, \ref{so31-dS2-0}) rewritten using the relevant $g_1/g_2$ scaling (see table \ref{table:dS5}) are
\beq
\begin{array}{c} dS_3\times H^2\\ SO(1,1)^{(2)}_\text{diag} \times SO(d,1)\\ d=2,3\end{array} \qquad\lf\{
\begin{array}{l}
\S_0 = c_1\, \zeta^{(2)}_{dS_3}\\\\
f_0 = c_2\, \lf(\zeta^{(2)}_{dS_3}\rr)^2 \\\\
g_0 = \frac{1}{2}\log \lf\{ c_3 \lf(\sqrt{1-2 a^2 g_1^2} -1 \rr) \dfrac{\lf(\zeta^{(2)}_{dS_3}\rr)^2}{g_1^2}\rr\}
\\\\
\zeta^{(2)}_{dS_3} = \lf[\dfrac{2 a^2 g_1^2-\sqrt{1-2 a^2 g_1^2}-1}{a^2 g_2^2}\rr]^{1/6}\end{array}\rr. \label{dS3-so112}
\eeq
\beq
\begin{array}{c} dS_3\times H^2\\ SO(1,1)^{(3)}_\text{diag} \times SO(2,1)\end{array} \qquad \lf\{
\begin{array}{l}\S_0= c_1 \,\zeta^{(3)}_{dS_3},\\
f_0= c_2\, g_1 \,\lf(\zeta^{(3)}_{dS_3}\rr)^2,\\
g_0= \dfrac{1}{2} \log \left(c_3 \left(\sqrt{1-3a^2 g_1^2}-1\right)\dfrac{\lf(\zeta^{(3)}_{dS_3}\rr)^2}{g_1^2}\right), 
\\\zeta^{(3)}_{dS_3}=\lf[{\dfrac{3a^2 g_1^2-\sqrt{1-3a^2 g_1^2}-1}{a^2 g_2^2}}\rr]^{1/6} \end{array}\rr. \label{dS3-so213}
\eeq
%%%%%%
\beq
\begin{array}{c} dS_2\times H^3\\ SO(1,1)^{(2)}_\text{diag} \times SO(3,1)\end{array} \qquad\lf\{
\begin{array}{l}
\S_0= c'_1 \zeta^{(2)}_{dS_2},\\\\
f_0= c'_2\, \, \lf(\zeta^{(2)}_{dS_2}\rr)^2\,\,g_1,\\
\\
g_0= \dfrac{1}{2} \log \left\{c'_3\left(\sqrt{1-\frac{1}{2} a^2 g_1^2}-1\right) \dfrac{\lf(\zeta^{(2)}_{dS_2}\rr)^2}{ g_1^2}\right\}
\\\\
\zeta^{(2)}_{dS_2} = \lf[\dfrac{\frac{1}{3}a^2 g_1^2-\sqrt{1-\frac{1}{2} a^2 g_1^2}-1}{a^2 g_2^2} \rr]^{1/6}
\end{array}\rr. \label{dS2-so312}
\eeq
$(c_1, c_2, c_3)$ and $(c'_1, c'_2, c'_3)$  are two sets of constants specific to each gauging and these are given in Table \ref{table:dS2-dS3}.
%%%%%%%%%%%%%%%%%%%%%%%%%%%%%%%%
\begin{table}[!htbp]
\centering
%\resizebox{\textwidth}{!}
\begin{tabular}{|c|c|c|}
\hline
 Gauge group & $\begin{array}{c} dS_3\times H^2\\ \{c_1, c_2, c_3\} \end{array}$ & $\begin{array}{c} dS_2\times H^3 \\ \{c'_1, c'_2, c'_3\} 
 \end{array}$\\
\hline &&\\
$SO(1,1)\times SO(2,1)$ & $\begin{array}{c}
\lf\{2^{1/6}, \,2^{-2/3}, \,2^{1/3}\rr\}\\
(\ref{all-dS3}) \end{array}$& - \\&&\\\hline&&\\
 $SO(1,1)\times SO(2,2)$ &$\begin{array}{c}\lf\{1, \,\dfrac{1}{2},\, 1 \rr\}\\
 (\ref{all-dS3}) \end{array}$&-  \\&&\\\hline&&\\
$SO(1,1)\times SO(3,1)$ & $\begin{array}{c}\lf\{\lf(\dfrac{2}{3}\rr)^{1/6},\, 12^{-1/3}, \,\lf(\dfrac{2}{3}\rr)^{1/3}\rr\}\\
 (\ref{all-dS3}) \end{array}$&   $\begin{array}{c}
\lf\{ \sqrt{2}, \, \sqrt{2}, \, 4 \rr\}\\
 (\ref{all-dS2}) \end{array}$\\&&\\\hline&&\\
$SO(1,1)\times SO(4,1)$ & $\begin{array}{c}\lf\{3^{-1/6},\, \dfrac{3^{-1/3}}{2},\, 3^{-1/3}\rr\}\\
(\ref{all-dS3}) \end{array}$& 
 $\begin{array}{c}
\lf\{ 2^{1/3},\, \dfrac{1}{\sqrt{2}},\, 2 \rr\}\\
  (\ref{all-dS2}) \end{array}$\\&&\\\hline &&\\
 $SO(1,1)\times SU(2,1)$ &$\begin{array}{c}\lf\{6^{-1/6},\, \dfrac{6^{-1/3}}{2},\, 6^{-1/3}\rr\}\\ (\ref{all-dS3}) \end{array}$& 
  $\begin{array}{c}
\lf\{ 2^{1/6},\, 2^{-1/6}, \, 2^{4/3}\rr\}\\
   (\ref{all-dS2})\end{array}$\\&&\\\hline&&\\
   %%%
 $SO(1,1)^{(2)}_\text{diag}\times SO(2,1)$ &  $\begin{array}{c} \lf\{ 2^{1/6},\, 2^{-1/6},\, 2^{-2/3}\rr\}\\ \ref{dS3-so112}\end{array}$& -\\&&\\\hline&&\\
 %%%
  $SO(1,1)^{(3)}_\text{diag}\times SO(2,1)$ & $\begin{array}{c} \lf\{ 2^{1/6},\, 2^{-2/3}\sqrt{3},\, \dfrac{2^{1/3}}{3}\rr\} \\ \ref{dS3-so213} \end{array}$ &- \\&&\\\hline&&\\
   $SO(1,1)^{(2)}_\text{diag}\times SO(3,1)$ & $\begin{array}{c} \lf\{ \lf(\dfrac{2}{3}\rr)^{1/6}, \, \,\lf(\dfrac{9}{8\sqrt{2}} \rr)^{1/3}, \,\,\lf( \dfrac{1}{12}\rr)^{1/3} \rr\} \\ \ref{dS3-so112}\end{array}$ &  $\begin{array}{c} \lf\{ 2^{1/2},\, 2,\, 2\rr\} \\ (\ref{dS2-so312}\end{array}$\\&&\\\hline
\end{tabular}
\caption{A summary of the fixed point $dS_2$ and $dS_3$ solutions in all the eight gauge groups with $dS_5$ vacuum of  matter-coupled $N=4$ five-dimensional gauged supergravity.}\label{table:dS2-dS3}
\end{table}
\\\\
Each of the cosmological  solutions interpolating between  $dS_{5-d}\times H_d$ ($d=2,3$) and $dS_5$ was plotted separately in  Figs. \ref{fig:SO21-dS3-cf}, \ref{fig:SO22-dS3-cf}, \ref{fig:SO31-dS3-cf}, \ref{fig:SO41-dS3-cf}, \ref{fig:SU21-dS3-cf} for $dS_3$ solutions, and  in Figs. \ref{fig:SO31-dS2-cf}, \ref{fig:SO41-dS2-cf}, \ref{fig:SU21-dS2-cf} for $dS_2$ solutions. 
To gain a better perspective of the solutions in all the gauged theories, we combine them together in Fig. \ref{fig:all-dS2-cf} and Fig. \ref{fig:all-dS3-cf} for $dS_2$ and $dS_3$ solutions, respectively. 
%%%%%%%%%%%%%%%%%%%%%%%%%%%%%%%%%%%%%%%%%%%%%%%%%%%%%%%%%%%
%%%%%%%%%%%%%%%%%%%%%%%%%%%%%%%%%%%%%%%%%%%%%%%%%%%%%%%%%%%
%%%%%%%%%%%%%%%%%%%%%%%%%%%%%%%%%%%%%%%%%%%%%%%%%%%%%%%%%%%
%%%%%%%%%%%%%%%%%%%%%%%%%%%%%%%%%%%%%%%%%%%%%%%%%%%%%%%%%%%
    \clearpage
    \newpage
    %\newpage
%%%%%%%%%%%%%%%%%%%%%%%%%%%%%%%%%%%%%%%%%%%%%%%%%%%%%%%%%%%
\section{First-order systems}\label{pseudo}
%%%%%%%%%%%%%%%%%%%%%%%%%%%%%%%%%%%%%%%%%%%%%%%%%%%%%%%%%%%
In this section we analyze how the cosmological solutions studied above and their corresponding $dS_5$ vacua cannot arise from the first-order systems of equations that solve the second-order field equations. We begin by asking the following question: Do there exist some sets of first order pseudo-BPS equations that solve the second order equations of motion given in (\ref{dS3eom}) and (\ref{dS2eom}) ?  If such these systems of pseudo-BPS first-order equations exist, there must be some pseudo-superpotentials $W$ that are contained in these first order equations to which the second order equations for $dS_5$ reduce \cite{ST06}, \cite{ST0610}, \cite{VR-07}. As established by the domain wall/cosmology correspondence \cite{ST0610}, for every domain wall solution arising from a scalar potential $V$ with the corresponding superpotential $W$, its cosmological counterpart can arise from an opposite scalar potential $-V$ with the corresponding pseudo-superpotential $i W$. It then follows that the pseudo-BPS equations in the $dS$ case are the  same BPS equations in the $AdS$ case. On a related note, there exist fake supersymmetric systems whose field equations can be reduced to first-order equations, containing a so-called fake superpotential $W$, that have nothing to do with the vanishing of fermionic supersymmetry variations \cite{SUGRA-fakeness}. It was shown in \cite{VR-12}, \cite{VR-16} that the fake superpotential $W$ can always be found in the Hamilton-Jacobi (HJ) formalism via the HJ characteristic function whose factorization contains $W$. 
\\
\indent While some of the features of the domain wall/cosmology correspondence, as highlighted above, apply to this case, 
there are some subtleties in this situation that should be carefully considered. The most important point to take note is that the scalar potentials for the $AdS$ and $dS$ solutions of five-dimensional $N=4$ supergravity arise from entirely different gauged theories and as such they have completely different forms. More specifically, $V_{dS} \neq -V_{AdS}$.  Consequently, the pseudo-superpotentials in the $dS$ case, if they exist, cannot be obtained from the superpotential in the $AdS$ case, i.e. $W_{dS} \neq i W_{AdS}$. Rather, the existence of this $W_{dS}$ must be checked using the specific forms of the various $V_{dS}$ constructed from the eight gaugings given in Table \ref{table:dS5-V}.
\\\indent
On the other hand, the fact that all $dS_5$ vacua of 5D $N=4$ supergravity are unstable indicates that no pseudosupersymmetry and no pseudosuperpotential exist that can allow for the construction of some first-order equations that solve \ref{dS2eom}, \ref{dS3eom}. Nonetheless, it is still instructive to understand definitively the 
exact way that characterizes this lack of relevant pseudosuperpotentials for the $dS_5$ vacua of 5D $N=4$ supergravity. To this end, we will first derive the first-order BPS equations that solve the second-order equations for $dS_5$ solutions and check how these fail to admit $dS_5$ solutions. To accomplish this, we proceed as follows.  First, we will specify the various BPS equations for the $AdS$ case that solve the second-order equations of motion for $AdS_5$ and its 1/2-BPS $AdS_3\times \S_2$ and 1/4-BPS $AdS_2\times \S_3$ solutions. The utility of this exercise lies in the fact since that the equations of motion for $AdS$ and $dS$ solutions are almost identical, the first-order equations for the $dS$ case that solve the $dS$ equations of motion are identical in form to the BPS equations. Since these first-order equations necessarily contain some pseudo-superpotential $W$ that is constrained to satisfy a relation between itself and the scalar potential $V$ in each gauged theory, we will determine whether there exist some suitable pseudo-superpotentials for the $dS$ case based on the explicit scalar potentials of the  eight $dS$ gauged theories. The final answer will turn out to be negative. 
\\\indent
To derive the relevant field and BPS equations for the $AdS$ case, we recall that there are three choices for an $AdS_5$-gauged theory with the gaugings $U(1)\times SU(2)$, $U(1)\times SO(3,1)$ and $U(1)\times SL(3,\mb R)$ \cite{AdS5-bs}. The common feature of these three theories lies in their common subgroup $U(1)\times SU(2)$ which must be a subgroup of the $SO(5)$ R-symmetry group of $N=4$ supergravity. In fact, this feature is a deciding factor for these gauged theories to admit $AdS_5$ solutions \cite{AdS5_N4}. In particular, we note that the $SU(2)$ common subgroup of the three gauge groups are generated by the generators
\beq
X_3, \qquad X_4, \qquad X_5
\eeq
corresponding to the particular embedding tensor component $f_{MNP} = f_{345}$ that must be present in all these three gauge groups \cite{AdS5-bs}. This fact will be useful later in our analysis when we discuss the gauge fields that are needed to obtain $AdS_{2,3}\times \S_{3,2}$ solutions.
\\\indent
For the purpose of this particular analysis, in either type of gauged theories, $AdS_5$ or $dS_5$, all the fields of $N=4$ supergravity can be truncated out except for the dilaton $\S$ and the metric. Consequently, the bosonic Lagrangian (\ref{L-full}) reduces to
\beq
e^{-1}\mc{L}&=&\frac{1}{2}R-\frac{3}{2}\Sigma^{-2}\pd_\mu \Sigma \,\pd^\mu \Sigma - V. \label{L-bare}
\eeq
In the analyses involving the $AdS_{2,3}\times H^{3,2}$ solutions, we will make use of the Lagrangian (\ref{L-red}) with the relevant gauge fields turned on. 
%%%%%%%%%%%%%%%%%%%%%%%%%%%%%%%%%%%%%%%%%%%%%%%%%%%%%%%%
\subsection{$AdS$ case}
%%%%%%%%%%%%%%%%%%%%%%%%%%%%%%%%%%%%%%%%%%%%%%%%%%%%%%%%
The ansatz for $AdS_5$  solution is
\beq
 ds^2 =dr^2 + e^{2f(r)}\lf[ -dt^2 +dz^2 +z^2\, d\Omega_2^2\rr], \label{adS5-a}
\eeq
with $d\Omega^2_2$ is the line element for $S^2$ and $H^2$ (\ref{O2}).  With the ansatz (\ref{adS5-a}), the equations of motion derived from (\ref{L-bare}) in an $AdS_5$ gauged theory are\footnote{Throughout the paper, we have used $\dot{\,}$ to denote derivatives with respect to $t$, here $^\prime$ will be used to denote derivatives with respect to $r$.}
\beq
0&=&3 f''+6 f'^2+\frac{3}{2} \frac{\Sigma '^2}{\Sigma ^2}+V(\Sigma),
\nonumber\\
0&=&6 f' \frac{\Sigma '}{\S}+\frac{3}{2} \frac{\Sigma ''}{\S}-\frac{3}{2} \frac{\Sigma '^2}{\Sigma^2 }-\frac{1}{2} \Sigma\,\partial_\Sigma V. \label{AdS5-eom}
\eeq
It can be easily verified that the following set of first-order equations can  solve the above set of second-order equations of motion (\ref{AdS5-eom}).
\begin{eqnarray}
\Sigma' &=&\Sigma^2 \,\partial_\Sigma\,W(\Sigma), \non
f'& =& -W(\Sigma) . \label{AdS5-bps}
\eeq
provided that $V$ must be related to $W$ as follows
\beq
 V(\Sigma) = \frac{3}{2}\Sigma^2 \left[\partial_\Sigma\,W(\Sigma)\right]^2 - 6 W^2(\Sigma). \label{vw}
\end{eqnarray}
Eqs. \ref{AdS5-bps} can be derived from setting to zero the supersymmetry transformations $\d\psi_\m^i$ and $\d\chi^i$ of the gravitini and dilatini, respectively. These equations are used to solve for holographic RG flow solutions \cite{AdS5-bs}.
\\\indent Given (\ref{AdS5-bps}), as long as there exists a superpotential $W$ satisfying (\ref{vw}), (\ref{AdS5-eom}) can reduce to (\ref{AdS5-bps}). For supersymmetric $AdS_5$ solutions, this is always the case, i.e. the superpotential $W$ always exists. As such, it is almost a trivial exercise to verify this. Recall that the scalar potential arising from any of the three gauged theories $U(1)\times SU(2), U(1)\times SO(3,1)$ and $U(1)\times SL(3, \mb R)$ is
\beq
V(\S) = -\frac{g^2_2}{2\S^2} + \sqrt{2}\,g_1\, g_2 \,\S, \label{Vu1su2}
\eeq
which admits the following $AdS_5$ vacuum at the origin of the scalar manifold (see \cite{AdS5-bs})
\beq
\S_0 = 1, \qquad g_1 = -\frac{1}{\sqrt{2}}g_2. \label{AdS5}
\eeq
The superpotential $W(\Sigma)$ is given by 
\beq
 W(\Sigma) = \frac{2g_2 - \sqrt{2}g_1 \Sigma^3}{6 \Sigma}.\label{ads5-w-spec}
\eeq
The superpotential (\ref{ads5-w-spec}) has the same $AdS_5$ critical point (\ref{AdS5}) as the scalar potential $V$ (\ref{Vu1su2}).
We note again that any of the three choices for the gauge group is completely equivalent and leads to the same results once all scalars from the vector multiplets are truncated out. 
\\\indent
Next, we will verify that the same superpotential $W$ (\ref{ads5-w-spec}) is part of the BPS equations that solve the second-order equations for the holographic RG flow solutions from an $AdS_{5-d}\times \S_d$ ($d=2,3)$ vacuum to an $AdS_5$ vacuum. For this type of solution, we make use of the Lagrangian (\ref{L-red}). These solutions were worked out in detail in \cite{AdS5-bs}, \cite{AdS5-bh} using the supersymmetric BPS equations obtained by setting to zero the supersymmetry variations of the fermionic fields. In this section, for the purpose of comparing with the $dS$ case later,  we will however derive their second order equations and show that these are solved by the same BPS equations, as derived in \cite{AdS5-bh}, \cite{AdS5-bs}, containing the superpotential $W$ (\ref{ads5-w-spec}).
\bei
\item $AdS_3\times \S_2$ case: The ansatz for this solution is
\beq
ds^2 = dr^2 + e^{2f(r)}\lf(-dt^2 + dx^2 +dy^2 \rr) + e^{2g(r)}d \Omega_2^2 \label{AdS3}
\eeq
where $d\Omega_2^2$ is given in (\ref{O2}). We also need to turn on an Abelian gauge field $U(1)$ with the same ansatz as (\ref{dS3-A-ans}). However, $M$ now must be a specific index along the $SO(5)$ symmetry directions, as opposed to the matter symmetry directions as in the $dS_5$ case. In particular, this gauge field can be chosen to correspond to the  $U(1)\subset SU(2)$ generated by $X_5$, and it reads explicitly
\beq
A^5 = \begin{dcases}a \cos \theta\, d\phi, & S^2\\ a\cosh\theta d\phi, & H^2 \end{dcases}\,.
\label{AdS3-A-ans}
\eeq
This gauge field is meant to implement a topological twist 
\beq
a\, g_2 = 1, \label{AdStwist}
\eeq
which cancels the non-trivial components of the spin connection coming from the $\S_2$ factor of the metric (\ref{AdS3}) in order to preserve some amount of supersymmetry.
The  equations of motion resulting from using the ansatze (\ref{AdS3}, \ref{dS3-A-ans}) in (\ref{L-red}) are
\beq
0 &=& -\lambda e^{-2 g}+\frac{1}{2} a^2 e^{-4 g} \S^2+f''+2 f' g'+f'^2+2 g''+3 g'^2+ \frac{3 \S'^2}{2 \S^2}+V(\S)\non
0&=& -\frac{1}{2} a^2 e^{-4 g} \S^2+2 f''+2 f' g'+3 f'^2+g''+g'^2+\frac{3 \S'^2}{2 \S^2}+V(\S)
\non
0 &=& \frac{1}{2} a^2 e^{-4 g} \S^2-3\frac{\S' \left( f'+ g'\right)}{ \S}-\frac{3 \S''}{2 \S}+\frac{3 \S'^2}{2 \S^2}+\frac{1}{2} \S\,V'(\S), \label{AdS3eom}
\eeq
with 
\beq
\lambda = \begin{dcases} 
+1, & \S_2 = S_2 \\-1, & \S_2 = H^2\end{dcases} \,\,.\label{lambdaAdS3}
\eeq
It can be seen that the three equations in (\ref{AdS3eom}) are almost identical to the three in (\ref{dS3eom}), save for the opposite signs in front of the non-derivative terms comprising the $V(\S), V'(\S)$ terms, the $\lambda e^{-2g}$ term, and the gauge field strength terms.
The set of second order equations (\ref{AdS3eom}) can be solved by the following set of first-order equations 
\beq
f'&=&  \frac{1}{3}\,\lambda\, a e^{-2 g} \Sigma+ W(\Sigma),
\nonumber\\
g'&=&  -\frac{2}{3}\,\lambda\, a e^{-2 g} \Sigma+W(\Sigma),\nonumber\\
\Sigma '&=&  \frac{1}{3}\,\lambda\, a e^{-2 g} \Sigma^2-\Sigma^2 \frac{\partial W(\Sigma)}{\partial \Sigma}, \label{AdS3-bps}
\eeq
subject to the conditions (\ref{vw}) and \ref{AdStwist}, where $W$ is still given by \ref{ads5-w-spec}. These first-order equations \ref{AdS3-bps} together with their $AdS_3\times H^3$ and associated RG flow solutions  can be found in \cite{AdS5-bs}.
Before moving on to the $AdS_2\times \S_3$ case, we remark that the set of second-order equations (\ref{AdS3eom}) admit to the same $AdS_3\times H^2$ fixed point solution and holographic RG flow interpolating between the aforementionned $AdS_3$ solution and the $AdS_5$ solution (\ref{AdS5}) as those found by solving the first-order systems (\ref{AdS3-bps}), as should be the case. 
%%%%%%%%%%%%%%%%%%%%%%%
\item $AdS_2\times \S_3$ case: The metric ansatz for this solution is
\beq
ds^2 = dr^2 + e^{2f(r)}\lf(-dt^2 + dz^2  \rr) + e^{2g(r)}d \Omega_3^2 \label{AdS2}
\eeq
where $d\Omega_3^2$ is given in (\ref{O3}). The $SO(3)$ non-Abelian gauge ansatz for this solution is given by (\ref{dS2-A-ans}) with the corresponding gauge fields given by (\ref{HMmn}). Note that the indices $M, N, P$ in (\ref{dS2-A-ans}) must now assume the specific values corresponding to the $SU(2)$ subgroup, generated by $X_3, X_4, X_5$, of the gauge groups $U(1)\times SU(2)$, $U(1)\times SO(3,1)$ and $U(1)\times SL(3,\mb R)$, which lie entirely along the $SO(5)\subset SO(5,n)$ R-symmetry directions. Explicitly,
\beq
\begin{dcases} A^3_\theta  = a \cos\psi, \,\,\, A^4_\phi = a \cos\theta,\,\,\, A^5_\phi = a\cos \psi \sin\theta, & \S_3 = S^3 
\\
A^3_\theta = a \cosh\psi, \,\,\, A^4_\phi = a \cos\theta, \,\,\, A^{5}_\phi = a \cosh\psi\, \sin\theta, & \S_3 = H^3 
 \end{dcases}
\label{AdS2-A-ans}
\eeq
This is in direct constrast to the $dS_2\times \S_3$ case where $M,N,P$ must be indices along the matter symmetry $SO(n)\subset SO(5,n)$ group of (\ref{sm}). The topological twist in this case is still given by (\ref{AdStwist}). 
\\\\
The resulting equations of motion from the ansatze (\ref{AdS2}, \ref{AdS2-A-ans}) are
\beq
0&=& -\lambda e^{-2 g}+\frac{1}{2} a^2 e^{-4 g} \S^2+g''+2 g'^2+\frac{\S'^2}{2 \S^2}+\frac{1}{3} V(\S)\non
0&=& -\lambda e^{-2 g}-\frac{1}{2} a^2 e^{-4 g} \S^2+f''+2 f' g'+f'^2+2 g''+3 g'^2+\frac{3 \S'^2}{2 \S^2}+V(\S)\non
0&=& a^2 e^{-4 g} \S^2-\S' \left(\frac{f'}{\S}+\frac{3 g'}{\S}\right)-\frac{\S''}{\S}+\frac{\S'^2}{\S^2}+\frac{1}{3} \S\, V'(\S), \label{AdS2eom}
\eeq
with 
\beq
\lambda = \begin{dcases} 1, & \S_3 = S^3\\ -1, & \S_3 = H^3\end{dcases}
\eeq
(\ref{AdS2eom}) are almost identical to (\ref{dS2eom}), except for the opposite signs in front of the $V(\S)$, $V'(\S)$, $\lambda e^{-2g}$, and gauge field strength terms (the non-derivative terms). The first-order equations that solve (\ref{AdS2eom}) are 
\beq
f'&=&-\lambda\, a e^{-2 g} \Sigma+W(\Sigma),
\nonumber\\
g'&=& \lambda\,a e^{-2 g} \Sigma + W(\Sigma),
\nonumber\\
\Sigma '&=& -\lambda\,a e^{-2 g} \Sigma^2-\Sigma^2 \frac{\partial W(\Sigma)}{\partial \Sigma}. \label{AdS2-bps}
\eeq
 subject to the conditions (\ref{vw}) and (\ref{AdStwist}), where $W$ is given by \ref{ads5-w-spec}.
The equations (\ref{AdS2-bps}) together with their $AdS_2\times H^3$ and associated RG flow solutions can be found in \cite{AdS5-bh}.
Finally, we note that the set of second-order equations (\ref{AdS2eom}) give rise to the same $AdS_2\times H^3$ fixed point and RG flow solutions as that found by solving the first-order systems (\ref{AdS2-bps}) as expected. 
\eni
Having done the analysis for the $AdS$ case, we remark that the same superpotential $W$, subject to the condition (\ref{vw}), appears in the BPS equations for the $AdS_5$ and $AdS_{2,3}\times \S^{3,2}$ cases. 
%%%%%%%%%%%%%%%%%%%%%%%%%%%%%%%%%%%%%%%%%%%%%%%%%%%%%%%%
\subsection{$dS$ case}
%%%%%%%%%%%%%%%%%%%%%%%%%%%%%%%%%%%%%%%%%%%%%%%%%%%%%%%%
The ansatz for $dS_5$  solution is
\beq
ds^2 =-dt^2 + e^{2f(t)}\lf[ dr^2 + dz^2 + z^2\, d\Omega_2^2\rr]. \label{dS5-a}
\eeq
With the ansatz (\ref{dS5-a}), the equations of motion derived from (\ref{L-bare}) in a $dS_5$ gauged theory are
\begin{eqnarray}
0&=& 3 \ddot f + 6 \dot f^2 + \frac{3}{2} \frac{\dot \Sigma ^2}{ \Sigma^2} - V(\Sigma ),
\nonumber\\
0&=& 6 \dot f \frac{\dot \Sigma}{\Sigma} +\frac{3}{2}\frac{\ddot \Sigma}{\Sigma} -\frac{3}{2}\frac{\dot \Sigma ^2}{ \Sigma^2}+\frac{1}{2} \Sigma\,  \partial_\Sigma V.  \label{dS5-eom}
\end{eqnarray}
These are almost identical to the equations of motion for the $AdS_5$ case (\ref{AdS5-eom}), except for the opposite signs in front of the potential terms $V(\S)$ and $V'(\S)$.
The following set of first-order equations can be verified to solve the above set of second-order equations of motion (\ref{dS5-eom})
\begin{eqnarray}
\dot\Sigma &=&\Sigma^2 \partial_\Sigma\,W(\Sigma), \non
\dot f &=& -W(\Sigma) \label{dS5-bps}
\end{eqnarray}
provided that there exists a $W(\Sigma)$ such that the scalar potential can be written as
\beq
V(\Sigma) = -\frac{3}{2}\Sigma^2 \left[\partial_\Sigma\,W(\Sigma)\right]^2 + 6 W^2(\Sigma).  \label{vw-dS}
\eeq
Comparing (\ref{AdS5-bps}) with (\ref{dS5-bps}), it is obvious that the two sets of first-order equations are  identical. The relation (\ref{vw-dS}) between the $dS$ scalar potential and its pseudosuperpotential $W$ has the same form but with opposite sign to that of the $AdS$ case (\ref{vw}). 
Given (\ref{dS5-bps}) and (\ref{vw-dS}), the construction of the first-order systems for the $dS_5$ case reduces to the determination of the pseudosuperpotential $W$ such that (\ref{vw-dS}) is satisfied. 
\\\indent
Unfortunately, with the specific form of $V(\Sigma)$ determined from the eight gaugings (see Table \ref{table:dS5-V}), there does not seem to exist a suitable $W(\Sigma)$ such that (\ref{vw-dS}) is satisfied. Instead, with $V$ as given in Table \ref{table:dS5-V}, we find the following pseudosuperpotentials that admit the same $dS_5$ critical points as the scalar potentials in  Table \ref{table:dS5-V}
\beq
W_\pm = \begin{dcases}
\frac{2g_2 \pm g_1 \S^3}{6\S}, & SO(1,1)\times SO(2,1) \\
\frac{2\sqrt{2}g_2 \pm g_1 \S^3}{6\S}, & SO(1,1)\times SO(2,2) \\
\frac{2 \sqrt{3}g_2 \pm g_1 \S^3}{6\S}, & SO(1,1)\times SO(3,1) \\
\frac{2 \sqrt{6}g_2 \pm g_1 \S^3}{6\S}, & SO(1,1)\times SO(4,1) \\
\frac{4 \sqrt{3}g_2 \pm g_1 \S^3}{6\S}, & SO(1,1)\times SU(2,1) \\
-\frac{2 g_2 \pm \sqrt{2} g_1 \S^3}{6\S}, & SO(1,1)^{(2)}_\text{diag}\times SO(2,1) \\
-\frac{2 g_2 \pm \sqrt{3} g_1 \S^3}{6\S}, & SO(1,1)^{(3)}_\text{diag}\times SO(2,1) \\
-\frac{\sqrt{3} g_2 \pm \sqrt{2} g_1 \S^3}{6\S}, & SO(1,1)^{(2)}_\text{diag}\times SO(3,1)
 \end{dcases} \label{dSW}
 \eeq
 In terms of $W_\pm$, the scalar potential $V$ (\ref{table:dS5-V}) can be written as
 \beq
 V = \frac{3}{2}\S^2\,\lf(\frac{\partial W_\pm}{\partial \S}\rr)^2 + 3 W_\pm^2.
 \eeq
 With the lack of a pseudo-superpotential that satisfies (\ref{vw-dS}), there does not exist a system of first-order pseudo-BPS equations for the $dS_5$ vacuum in the eight $dS_5$ gauged theories of $N=4$ 5D supergravity. 
 By a direct analogy with the $AdS_{2,3}\times \S_{3,2}$ cases analyzed in detail above, if the first-order systems for the $dS_{2,3}\times \S_{3,2}$ cases existed, they would be of the same form as (\ref{AdS3-bps}, \ref{AdS2-bps}) with the pseudosuperpotential $W$ satisfying (\ref{vw-dS}).  The non-existence of a $W$ satisfying (\ref{vw-dS}), therefore, means that no first-order systems can be constructed for the $dS_{2,3}\times \S_{3,2}$ cases that solve \ref{dS3eom}, \ref{dS2eom}. 
 \\\indent
 As previously mentioned, this conclusion that no pseudosuperpotentials and subsequently no first-order equations exist for the $dS_5$ and $dS_{5-d}\times \S_d$ solutions of 5D $N=4$ supergravity could have been reached without doing any of the work above. In the same manner that supersymmetry guarantees the stability of supersymmetric solutions, pseudosupersymmetry ensures the stability of pseudosupersymmetric solutions. Unstable solutions are necessarily non-pseudosupersymmetric. 
Furthermore, the lack of a suitable pseudosuperpotential $W$ for a $dS$ solution can also be due to the fact that there exist some mass values in the mass spectrum of this $dS$ solution that violate the Breitenlohner-Freedman (BF) bound,\footnote{I would like to thank Thomas Van Riet for pointing this out.} which, in the case of cosmology, reads
\beq
m^2 L^2 \leq \frac{(D-1)^2}{4} \label{BFb}
\eeq
where $D$ is the spacetime dimensions, and $L$ is the radius of $dS$ space. 
With $D=5$, $L$ is given by $\sqrt{6/V_0}$, and the bound (\ref{BFb}) becomes
\beq
m^2 L^2 \leq 4. \label{BFb5d}
\eeq
 It can be verified from the mass spectra  given in \cite{dS5} that there are mass values violating the bound (\ref{BFb5d}) for all known $dS_5$ solutions of five-dimensional $N=4$ supergravity. 
%%%%%%%%%%%%%%%%%%%%%%%%%%%%%%%%%%%%%%%%%%%%%%%%%%%%%%%%%%%
\section{Conclusions}\label{concl}
%%%%%%%%%%%%%%%%%%%%%%%%%%%%%%%%%%%%%%%%%%%%%%%%%%%%%%%%%%%
In this work, we have studied the time-dependent cosmological solutions between a $dS_{5-d}\times H^d$ with $d=2,3$, spacetime in the infinite past to a $dS_5$ spacetime in the infinite future from five-dimensional $N=4$ matter-coupled supergravity.  It is worth reiterating that our motivation for this study is completely decoupled from dS/CFT holography. In carrying out this work, our sole aim was to find out whether there exist fixed-point solutions of the form $dS_{5-d}\times \S_d$ with $\S_d = S^d, H^d$ where $d=2,3$, and corresponding interpolating solutions between these at $t\ra -\infty$ and the $dS_5$ solutions, found in \cite{dS5}, at $t\ra \infty$. As such, our starting point for the derivation of these solutions is the eight gauged theories whose gauge groups are capable of admitting $dS_5$ solutions.
We note that the gauge fields that are required to support the $\S_d$ factor correspond to the compact generators embedded entirely the matter symmetry group of the global symmetry group. While $dS_3\times H^2$ solutions are present in all eight theories due to the fact that only an Abelian factor $SO(2)$ is required, $dS_2\times H^3$ solutions are only possible in those gauged theories whose gauge groups are large enough to contain an $SO(3)$ factor required to turn on the non-Abelian $SO(3)$ gauge fields. To the best of our knowledge, these are the first  cosmological solutions arising from half-maximal supergravity. Since all known $dS_5$ vacua of 5D $N=4$ are unstable, it is understood that these and their associated cosmological solutions cannot be pseudosupersymmetric. Consequently, these solutions can only arise from the second-order equations of motion.  We explicitly verified this by analyzing how the cosmological solutions and their corresponding $dS_5$ vacua fail to arise from the first-order systems of equations that solve the second-order equations of motion. Relying on the parallel analysis for the $AdS$ case, we derived the form of the first-order equations for the $dS$ case and verified that no suitable pseudosuperpotentials exist that could allow for the existence of the constructed first-order systems.  
\\\indent
A rather puzzling feature of the solutions found here is the fact that the square of the product of the gauge flux $a$ and gauge coupling constant $g_2$ must be negative. This is reminiscent of the case for the cosmological solutions of the $dS$ supergravities, with the wrong sign for the  kinectic gauge terms, arising from the dimensional reduction of the exotic $M^*$ and IIB$^*$ theories \cite{Lu-03}. Given that these solutions are derived from a conventional supergravity with the correct sign for the kinetic term of the gauge fields, it was rather unexpected to observe this feature. Two alternatives now arise to deal with this. The first one is to disregard these solutions altogether on the basis of the pathological nature of the imaginary gauge flux required to obtain real solutions, which is equivalent to flipping the sign of the kinectic terms for the gauge fields.  The second one is to interpret these solutions as providing an additional characterization of the dS vacuum structure of the eight half-maximal gauged supergravity theories constructed from the eight gaugings of \cite{dS5}.  With our viewpoint on this matter being that of the second alternative, it would be interesting to us to understand the implications of this. 
\\\indent
 Given the similarity between the 5D and 4D $N=4$ supergravities, the analysis done in this work for the 5D theory can be performed in the more complex 4D case. 
We recall that the $dS$ theories constructed here are based on those gauge groups whose compact subgroups must be embedded entirely in the matter directions of the scalar manifold. These gauge groups can only give rise to $dS$ solutions without the possibility of giving rise to $AdS$ solutions. In \cite{dS4}, these gaugings are classified to be of the second type, in constrast to the first type of $dS$ gaugings that can give rise to both $dS$ and $AdS$ solutions with different ratios of gauge couplings. It was established in \cite{dS5} that in five dimensions, $dS$ gaugings of the first type do not exist, while in four dimensions, $dS$ gaugings of both type do, as shown in \cite{dS4}. Accordingly, it would be interesting to extend the analyses here to the four-dimensional, half-maximal case with both types of $dS$ gaugings, to firstly derive cosmological solutions, and to secondly characterize how these 4D solutions fail to arise from the relevant first-order equations that solve the second-order equations of motion. 
These issues were investigated and reported in the companion work \cite{dS4-cosmo}. While the full details can be found there, we would like to briefly comment on the obtained results in four dimensions. Unfortunately, similar pathologies to the 5D case persist in the 4D cosmological solutions derived in \cite{dS4-cosmo}, both in the first and second types of gaugings. This is probably related to the fact that all four-dimensional gaugings can be derived from the five-dimensional ones by dimensional reduction as pointed out in \cite{dS4}. These undesirable features of the 4D and 5D $dS$ solutions imply that the corresponding gaugings from which the gauged supergravity theories are constructed are not physically interesting. However, these gaugings only exhaust the list of all possible semisimple groups (both in four and five dimensions), leaving open the possibility of non-semisimple gaugings being capable of admitting stable $dS$ solutions. If these exist, they will possibly lead to physically relevant gauged supergravity theories with potentially physically interesting $dS$ vacuum structures.
\\\\
\textbf{Acknowledgements}: I thank Thomas Van Riet for a helpful discussion regarding first-order systems. 
The author is supported by the grants C‐144‐000‐207‐532 and C‐141‐000‐777‐532 for postdoctoral research. 
%\clearpage
%%%%%%%%%%%%%%%%%%%%%%%%%%%%%%%%%%%%%%%%%%%%%%%%%%%%%%%%%

\end{document}